\documentclass[%
 reprint,
 amsmath,amssymb,
 aps,pre
]{revtex4-2}

\usepackage{graphicx}
\usepackage{dcolumn}
\usepackage[caption=false]{subfig}
\usepackage{bm}
\usepackage{float}
\usepackage{mathtools,amsmath,amssymb}
\usepackage{hyperref}

\begin{document}

\preprint{}

\title{Critical properties of Heider balance on multiplex networks}

\author{Krishnadas Mohandas}
\email{krishnadas.mohandas.dokt@pw.edu.pl}
 \author{Krzysztof Suchecki}
 \email{krzysztof.suchecki@pw.edu.pl}
\author{Janusz A. Holyst}
 \email{janusz.holyst@pw.edu.pl}
\affiliation{
Faculty of Physics, Warsaw University of Science and Technology\\
Koszykowa 75, PL-00-662 Warsaw, Poland
}

\date{\today}

\begin{abstract}
Heider's structural balance theory has proven invaluable in comprehending the dynamics of social groups characterized by both friendly and hostile relationships. Since people's relations are rarely single-faceted, we investigate Heider balance dynamics on a multiplex network, consisting of several copies of the same agent displaying correlated relations at different layers building the multiplex. 
Intralayer interactions in our model adhere to Heider dynamics, while interlayer correlations stem from Ising interactions, with the heat bath dynamics of link signs. 
The investigations uncover a multifaceted system with a diverse equilibrium landscape contingent on the coexistence of distinct phases across layers. We observe that starting from a paradise state with positive links in all layers, an increase in temperature triggers a discontinuous transition to a disordered state akin to single-layer scenarios. The critical temperature surpasses that of the single-layer case, a fact verified through extended mean-field analysis and agent-based simulations.
Furthermore, the scenario shifts when one layer exhibits a two-clique configuration instead of a paradise state. This change introduces additional transitions: synchronization of inter-layer relations and a transition to the disorder, appearing at a different, lower temperature compared to matching paradise states. This exploration shows the intricate interplay of Heider balance and multiplex interactions.
\end{abstract}


\maketitle

\section{\label{sec:level1}Introduction}

Multiplex structures \cite{kivela2014multilayer} find widespread application in characterizing diverse systems such as social groups, transportation networks, and biological frameworks, including protein-protein interaction systems \cite{murase2014multilayer,sole2016congestion,szell2010multirelational,boccaletti2014structure,hosni2020minimizing,singh2008global,gomez2012evolution}.
Humans are inherently complex beings, and when they form interconnected groups bound by relationships, the intricacy only deepens.
Individuals frequently maintain a plethora of relationships, varying in nature from familial and professional ties to friendships and online connections.
A multiplex network representation is invaluable in addressing this diversity of relations \cite{szell2010multirelational,salehi2015spreading,arinik2020partitioning}.
Through distinct layers or edge types, this representation delineates the assorted interaction types.
Incorporating the multiplex essence of these relationships provides an enriched comprehension of the system's dynamics.
In certain instances, this approach unveils hitherto undiscovered mechanisms or phenomena that remain concealed within simple, aggregated network models \cite{buldyrev2010cascading,atkisson2020understanding}.

When examining human interactions, a fundamental categorization involves distinguishing between friendly and hostile relations.
Within a group of individuals, these relations can be effectively portrayed using signed networks \cite{leskovec2010signed}.
Positive links denote friendly relations between nodes representing individuals, while negative links signify hostile relations.
In this context, Heider introduced the Structural Balance Theory (SBT) to social psychology as a means to delineate the underlying tensions within such networks \cite{heider1946attitudes}. Central to Heider's theorem is the principle that a signed network attains balance when all triads exhibit either three positive relations or one positive and two negative relations \cite{antal2006social,antal2005dynamics}.

Balance theory posits that networks characterized by friend-or-foe relationships evolve towards a state of greater equilibrium \cite{srinivasan2011local}. Notably, Cartwright and Harary demonstrated that, beyond an all-positive "paradise" state, a complete graph is balanced if agents can be segregated into two groups, featuring solely positive ties within each group and exclusively negative ties bridging the groups \cite{cartwright1956structural,marvel2011continuous}.

Recent studies have extended the concept of structural balance to the realm of statistical physics, establishing a parallel between the tension of imbalanced triads and energy excitation above the system's ground state \cite{rabbani2019mean,malarz2021comment,shojaei2019phase}.
This framework enables the incorporation of uncertainty regarding individual actions as thermal noise, quantifiable through temperature.
By adopting this approach, the need for specific microscopic dynamical rules governing changes in relations is obviated.
Furthermore, it harnesses established statistical methods from physics to describe and predict system behavior.

Several models have been developed, grounded in the principles of SBT, aiming to elucidate human behavior within social networks \cite{summers2013active,xia2015structural,yang2022promotive,facchetti2011computing,doreian2015structural}. However, the exploration of structural balance within the realm of multiplex networks remains limited to a handful of investigations \cite{gorski_destructive_2017,kundu2022balance,cozzo2015structure,burghardt2018imbalance,kulakowski2005heider}.

Within this study, we embark on an extension of the Heider Balance (HB) concept to the domain of multiplex networks, wherein nodes (agents) are linked by various kinds of relations corresponding to different network layers.
Our premise is rooted in the idea that individuals exhibit a predilection for consistency; if they share a friendly link within a work context, this amiability is likely to extend beyond the workplace.
In cases where this consistency wavers, we posit the emergence of internal tension.
To capture this concept, we introduce a coupling mechanism that fosters similarity among {\it relations} linking {\it the same} pair of individuals, akin to the coupling between spins in the Ising model of ferromagnets.

Our objective is to understand the impact of layer-to-layer coupling on the signed relation configurations across diverse layers.
We start by considering a duplex network configuration, where intralayer relations in each layer form a complete graph, and interlayer Ising interactions are exclusive to the same link replicas.
Subsequently, we generalize this model to encompass an arbitrary number of layers.
The dynamics of relation signs within each layer are governed by Heider Structural Balance dynamics.
Conversely, Ising dynamics dictate the interlayer coupling of relation signs concerning the same pairs of individuals. 

The forthcoming section extends the classical mean-field theory of HB to the context of multiplex networks.
Our theoretical approach correctly predicts the occurrence of a discontinuous order-disorder transition induced by thermal noise, including both duplex and multiplex network cases.
The subsequent section employs Monte Carlo simulations to validate the analytical findings.
Further investigation includes a more thorough overview of HB dynamics within a duplex network, where layers can exist in different states, not described by the analytical approach.
The dynamics of such a system encompass a temperature-driven transition toward synchronization between layers in different states and a subsequent transition toward disorder.

\section{Model}
Our model represents people as agents placed in vertices of a multiplex network, with dynamical relations between them represented by signed edges.
Edges belonging to a specific layer correspond to some type or context of relation (such as workplace or private relations), while the same vertices exist across all layers.
The vertices representing agents have no dynamical attributes, while edge signs obey Heider structural balance related to interactions within each layer and Ising-type dynamics related to interactions between layers to take into account positive correlations between different relations for the same pair of agents .\\

\begin{figure}
    \centering
    \includegraphics[width=\linewidth]{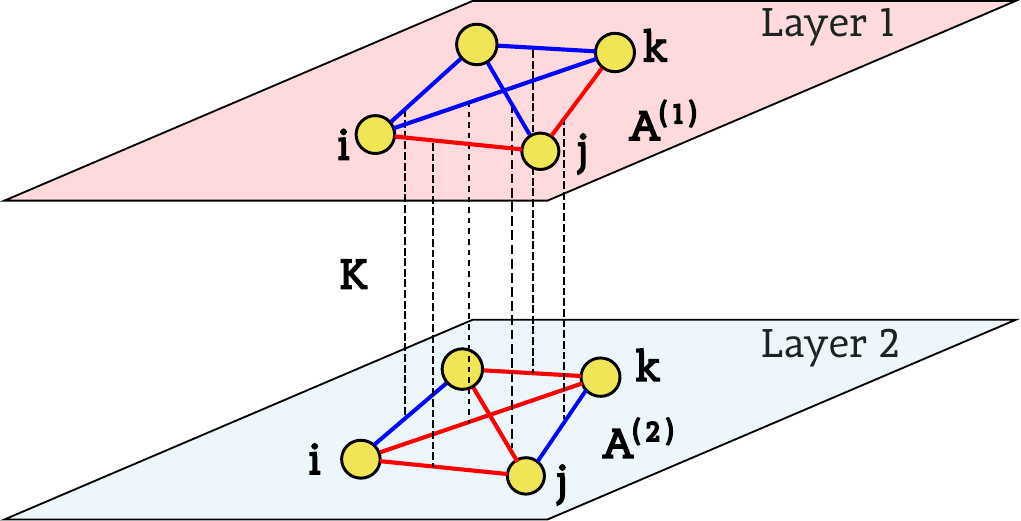} 
	\caption{An example of interactions between edge signs (blue for $+1$, red for $-1$) on a duplex network. Edge sign changes are driven by intralayer interactions according to Heider structural balance (with strength $A^{(1)}$ or $A^{(2)}$ respectively) as well as interlayer coupling of Ising nature between the edges, with strength $K$. The signs of relations evolve over time, according to the heat bath dynamics with energy as expressed by Eq.\ref{hamiltonian} and given temperature $T$.}
    \label{fig:bilayer}
\end{figure}

Let us consider a multiplex network with $N$ nodes and $L$ layers.
For simplicity, we assume that every pair of agents has a relation between them, meaning that the topology of each layer is a complete graph.
Each pair of agents $i,j$ has therefore $L$ signed relations $x_{ij}^{(\alpha)}=\pm 1$ between them, where $i$ and $j$ indicate agents and $\alpha=1,2,\ldots,L$ indicates the layer.
The edge signs $x_{ij}^{(\alpha)}$ are dynamic variables and can change over time.
We assume that each imbalanced triad $(ij,jk,ki)$ of relations causes a certain tension that the agents $i,j,k$ involved try to relax.
This is represented as a specific energy associated with the triad: $-A^{(\alpha)}$ if the triad is balanced and $+A^{(\alpha)}$ if it's not, according to Heider structural balance theory.
Energy $A^{(\alpha)}>0$ represents the strength of the Heider coupling between link signs within each layer.
Similarly, a discrepancy between relations in different layers between the same pair of agents $i,j$ causes a tension, also represented as energy: $-K$ if the relations are the same and $+K$ if they are different.
Energy $K>0$ corresponds to the ferromagnetic Ising model coupling between link signs across layers.
Fig. \ref{fig:bilayer} illustrates the model for a duplex network.

The Hamiltonian $H$ of the system, or its energy $E$ is
\begin{multline}
H = -\sum_{\alpha=1}^{L} A^{(\alpha)} \sum_{i>j>k} x_{ij}^{(\alpha)} x_{jk}^{(\alpha)} x_{ki}^{(\alpha)}\label{hamiltonian}\\
-K \sum_{\alpha>\beta} \sum_{i>j} x_{ij}^{(\alpha)} x_{ij}^{(\beta)}
\end{multline}
The state of the system will change according to the Hamiltonian in the presence of temperature $T$, which represents the uncertainty of relations or tolerance towards imbalanced relations in the social systems modeled.\\
The model has degenerate ground states, where any division of vertices into two groups is a ground state, if all intra-group edges in all layers are positive and all inter-group edges are negative \cite{antal2006social}.
We will call this state a \emph{two-clique state}.
This also includes the so-called \emph{paradise state}, where one group contains all vertices and the other no vertices so that all relations are positive.
Due to thermal fluctuations, at any temperature $T>0$ the system settles into an equilibrium different than the ground state and, in fact, is multistable, with the exact state reached via a dynamical process depending not only on parameters but also on the initial conditions.

\section{Analytic Approach}

\subsection{Mean field approximation for a duplex network}
\label{sec:meanfield}
Let us consider a duplex network, a particular case of a system described by Hamiltonian (\ref{hamiltonian}) corresponding to  $L=2$ where each of both link layers has the structure of the complete graph. A mean-field approach for a single network has been proposed in \cite{malarz_mean-field_2022}, where each single link $x_{ij}$ interacts with the mean-field proportional to $\left\langle x\right\rangle^2$ as a result of the presence of this link in all $N-2$ triads.
Here $\left\langle\dots\right\rangle$ indicates an ensemble average over all possible states.
This method works accurately for a single-layer complete graph with $N>>1$ since each individual link interacts with a significant fraction of all other links, which is accurately captured by the mean-field approach.
Extending this approach to multiplex networks is not as straightforward as it may seem at first glance.
A naive inclusion of the second layer influence as an Ising mean field would represent each edge having an Ising-type interaction with all the edges of the second layer, instead of only the single edge it actually has interaction with.
This can drastically alter the behavior of the system, especially for strong interlayer interactions.\\
In order to tackle this for a duplex network, a mean-field approach is employed with a pair of coupled links $\vec{x}_{ij}=[x_{ij}^{(1)}, x_{ij}^{(2)}]$ instead of a single link $x_{ij}^{(1)}$ as our elementary subsystem that will be interacting with the mean field.
Our pair $\vec{x}_{ij}$ will be experiencing interlayer interaction as internal energy of the state $\vec{x}_{ij}$ and intralayer interactions (in both layers) as the interaction of $\vec{x}_{ij}$ with a 2-dimensional mean field proportional to $[\langle x^{(1)} 
\rangle ^2,\langle x^{(2)}\rangle ^2]$.
Since we assume the system dynamics is expressed by state energy and thermal equilibrium in a specific temperature $T$, the probabilities of states $\vec{x}_{ij}$ of this pair are described by the canonical ensemble
\begin{equation}
    P(\vec{x}_{ij})= \frac{\exp(- E(\vec{x_{ij}})/T)}{\sum_{\vec{x}_{mn}}\exp(- E(\vec{x}_{mn})/T)} \label{eq:canonical}
\end{equation}
where
\begin{multline}
    E(\vec{x}_{ij})= -A^{(1)} \sum_{k\neq i,j}  x_{ij}^{(1)} x_{jk}^{(1)}x_{ik}^{(1)} -A^{(2)} \sum_{k\neq i,j}  x_{ij}^{(2)} x_{jk}^{(2)}x_{ik}^{(2)}\\
    -K x_{ij}^{(1)}x_{ij}^{(2)}
\end{multline}
We will assume the same interaction parameters for all the layers (i.e. $A^{(1)}=A^{(2)}=A$).
The expected value of $\left\langle \vec{x}_{ij} \right\rangle$ can be written as,
\begin{equation}\label{eq:meanx}
    \left\langle \vec{x}_{ij} \right\rangle =\sum_{\vec{x_{ij}}} P(\vec{x_{ij}}) \vec{x}_{ij}
\end{equation}
Following the mean-field method, we assume that the link sign $x_{ij}^{(\alpha)}$ interacts not with specific link sign product $x_{jk}^{(\alpha)} x_{ki}^{(\alpha)}$ but with a mean field $(x^{(\alpha)})^2$, with the mean link signs $x^{(\alpha)}$ that will be noted as
\begin{equation}
    \vec{x} \equiv [x^{(1)},x^{(2)}]
\end{equation}
We will call this mean link sign value a \emph{polarization} of the network (if considering the entire vector) or of the given layer (if considering specific components $x^{(\alpha)}$ of it).
This allows us to write energy of a given state $\vec{x}_{ij}$ as
\begin{multline}
    E(\vec{x}_{ij})= -A(N-2) x_{ij}^{(1)} {(x^{(1)})}^2 -A (N-2)x_{ij}^{(2)} {(x^{(2)})}^2 \label{eq:energyxij}\\
    -K x_{ij}^{(1)}x_{ij}^{(2)}  
\end{multline}

\begin{figure}[H]
    \includegraphics[width=\linewidth]{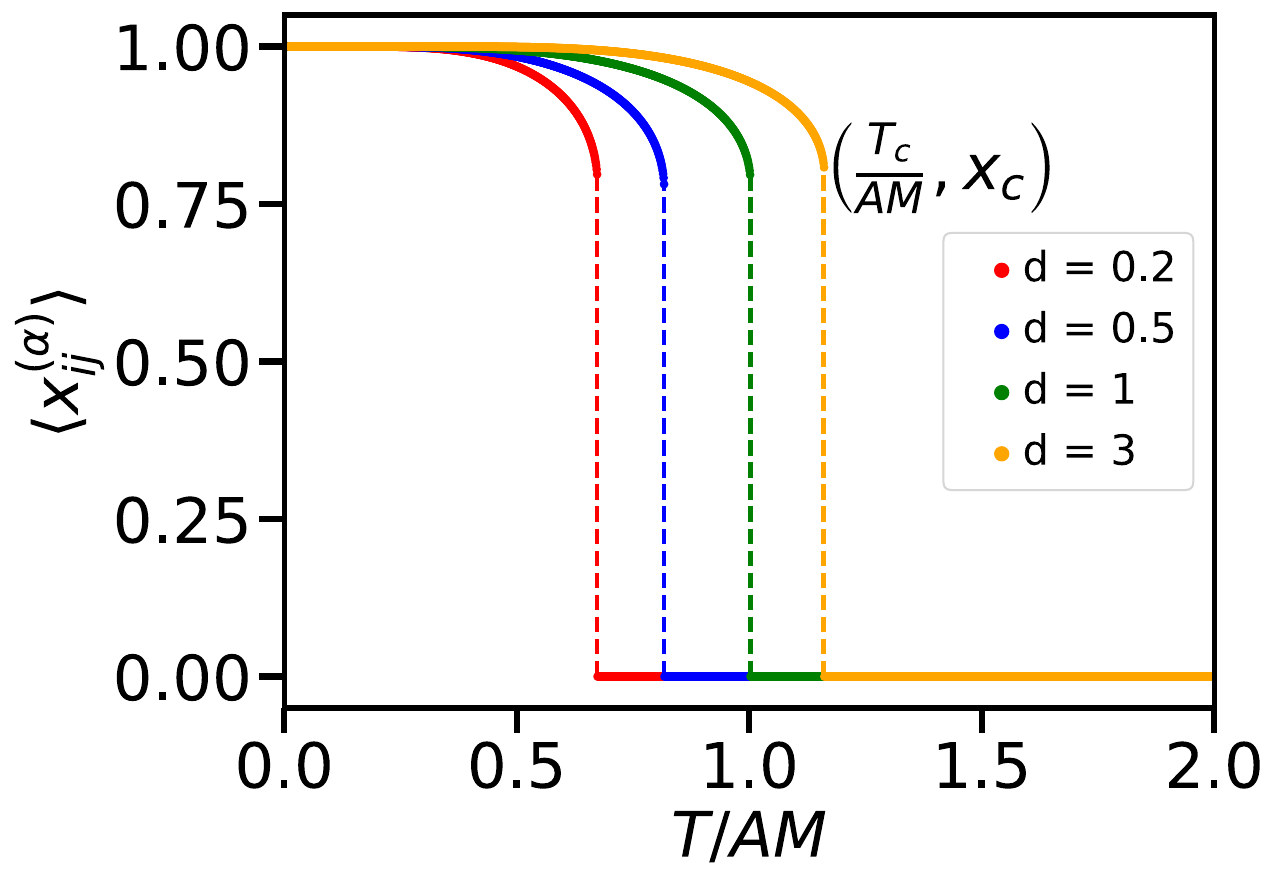} 
    \caption{Mean field solution for average link polarization $x^{(1)}$ for various interlayer coupling strength. The coupled layers undergo a discontinuous transition at a critical temperature $T_c$ that increases with coupling strength.}
    \label{fig:MF_2L}
\end{figure}

Combining (\ref{eq:energyxij}) and (\ref{eq:canonical}) into (\ref{eq:meanx}) allows us to calculate the expected value of link signs $\langle \vec{x}_{ij} \rangle$ in considered pair $ij$ (see Appendix \ref{App1}).
Assuming that the calculated $\langle \vec{x}_{ij} \rangle$ is the same as the $\vec{x}$ in the mean-field, meaning $\langle \vec{x}_{ij} \rangle = \vec{x}$, allows us to write a set of self-consistent equations for mean polarization $\vec{x}$

\begin{equation}
\begin{aligned}
x^{(1)}=f_1(x^{(1)},x^{(2)}) \label{eq:2dMFSolution}\\
x^{(2)}=f_2(x^{(1)},x^{(2)})
\end{aligned}
\end{equation}
where
\begin{multline}
f_{\alpha}(x^{(1)},x^{(2)})=\\
\frac{e^{2d}\sinh (a[{(x^{(1)})}^2\!\mathbb{+}{(x^{(2)})}^2]) \mathbb{+}\sinh (a(-1)^{\alpha}[{(x^{(2)})}^2\!\mathbb{-}{(x^{(1)})}^2])}{e^{2d}\cosh (a[{(x^{(1)})}^2+{(x^{(2)})}^2])+\cosh (a[{(x^{(1)})}^2-{(x^{(2)})}^2])}
\end{multline}
with $\alpha \in \{1,2\}$. The parameters $a=\frac{AM}{T}$ and $d=\frac{K}{AM}$ are re-scaled intralayer and interlayer interaction strength with $M=N-2$ being the number of triads each edge belongs to.\\
Fig. \ref{fig:MF_2L} shows the numerical solution of average link polarization based on mean-field solution (solution of Eq.\ref{eq:2dMFSolution}) for different values of rescaled interlayer coupling strength $d$.

We observe from Fig.\ref{fig:MF_2L} that by increasing temperature $T$, the mean polarization continuously decreases to a point $x_c$ when it jumps to zero. At this point, a first-order transition is observed at a critical temperature $T_c$. The values $T_c$ and $x_c$ depend on the coupling strength $K$ between the layers. Increasing the coupling strength increases $T_c$, asymptotically approaching a certain saturated $T_c$ value (as seen in Fig. \ref{fig:MF_ABMcomparison}).
Note that the critical temperature $T_c$ represents the temperature where a stable polarized state $(\vec{x} \neq 0)$ disappears.
Since the unpolarized state is always stable in the mean-field approach, this first-order transition is always from a polarized to an unpolarized state, the reverse transition does not take place and an unpolarized system will remain unpolarized even if the temperature is reduced below $T_c$.
The actual behavior of the system, outside the mean-field approach, is more complex (see Sec. \ref{sec:mismatched}).\\
To tackle the problem analytically, one can treat Eq.\ref{eq:2dMFSolution} as recurrence equations describing how the average link polarizations $[x^{(1)},x^{(2)}]$ evolve when influenced by the mean field and its past state.
This allows us to consider the issue of multistability that our system exhibits and how it could change between these states, instead of being limited to just finding equilibria.
The dynamical equations are
\begin{equation}
\begin{aligned}
x^{(1)}(t+1) = f_1(x^{(1)}(t),x^{(2)}(t)) \label{eq:2dmap}\\
x^{(2)}(t+1) = f_2(x^{(1)}(t),x^{(2)}(t))
\end{aligned}
\end{equation}
where $f_{\alpha}(x^{(1)}(t),x^{(2)}(t))$ are right-hand sides of Eq. \ref{eq:2dMFSolution} taken at time $t$.
Then, the fixed points of the map, Eq. \ref{eq:2dmap} are solutions of the implicit equation \ref{eq:2dMFSolution}.
It is worth noting that $\vec{x}_0=[0,0]$ is always a stable solution of Eq. \ref{eq:2dmap}, that for lower temperatures can co-exist with another stable solution.
When considering the behavior of the system, we are looking at the critical temperature $T_c$ where the non-zero solution disappears and the stable state of the system starting in paradise discontinuously changes from $x_c$ to $0$.\\
The values $T_c$ and $x_c$ can be received from a pair of transcendental algebraic relations (Eq.\ref{eq:transcendentalEqn} and Eq.\ref{eq:CriticalTemp}) that describe the fixed point and its Jacobian matrix at the point where largest eigenvalue of the Jacobian crosses $1$ (the critical point). Detailed calculations are shown in Appendix \ref{App1}. 
The relation between $T_c$ and $x_c$ is given by auxiliary variable $z=e^{\frac{AM}{T_c}x_c^2}$ that obeys
\begin{equation}\label{eq:transcendentalEqn}
    8\ln z = \frac{(z^4-1)(z^4D^2+2z^2+D^2)}{z^2(z^4+2z^2D^2+1)}
\end{equation}
where $D=e^{d}$.
Solving this equation numerically allows to find $z$ satisfying it, which depends only on $D$, and thus indirectly on the ratio of interlayer coupling $K$ and intralayer coupling $A$ (Fig. \ref{fig:solutionz} in Appendix shows the general shape of this solution).\\
Given the $z$ value, it's possible to find critical polarization $x_c(D,z)$ and the temperature $T_c(x_c,z)$.
\begin{equation}\label{eq:CriticalTemp}
     \frac{T_c}{A M}=\frac{x_c^2}{\ln z}=\left(\frac{1}{\ln{z}}\right)\left( \frac{D^2(z^4-1)}{D^2(z^4+1)+2z^2} \right)^2
\end{equation}
The dependence of this critical temperature on coupling can be seen in Fig. \ref{fig:MF_ABMcomparison}.
For $D=1$ (non-interacting layers), the approach is effectively reduced to one layer case and Eq.(\ref{eq:CriticalTemp}) gives the critical temperature of a single-layer network.

\subsection{Generalization to higher-order multiplex network}
In order to study multilayer effects, we generalize the duplex network considered in Sec. \ref{sec:meanfield} in the following way. We consider $L$ layers of the same set of nodes and assume that each layer corresponds to a different type of relationship or communication context. Edges in each layer interact with corresponding edges in all other layers since different relationships of the same pair of agents are correlated. For simplicity, the same parameters $A^{(\alpha)}=A$ are used for all the layers and the coupling strength between each pair of layers is the same $K$. 
The energy of all $L$ interactions between a given pair of agents $ij$ is
\begin{equation}
    E(\vec{x}_{ij})= -A\left( \sum_k^{M_{ij}}\sum_{\alpha=1}^L   x_{ij}^{(\alpha)}x^{(\alpha)}_{jk}x^{(\alpha)}_{ki} \right)-K\sum_{\alpha>\beta} x_{ij}^{(\alpha)} x_{ij}^{(\beta)}
    \label{Eq:energy_raw_multi}
\end{equation}
The approach used to analyze duplex networks can be used for a higher number of layers.
For $L=3$, using Eq.(\ref{eq:meanx}) with $\vec{x}_{ij}=(x_{ij}^{(1)}, x_{ij}^{(2)}, x_{ij}^{(3)})$ and energy (\ref{Eq:energy_raw_multi}) for 3-layered network, and following the same methodology as for duplex network, we can write set of self-consistent equations for $\vec{x}=(x^{(1)}, x^{(2)}, x^{(3)})$.
Using an analogous method as for duplex network (see Appendix), we arrive at the transcendental equation for an auxiliary variable $z=z(D)=e^{\frac{AM}{T_c} x_c^2}$
\begin{equation}\label{eq:transcendentalL3}
    8\ln\left(z\right)=\dfrac{\left(D^4z^6+z^4-z^2-D^4\right)\left(D^4z^6+3z^4+3z^2+D^4\right)}{z^2\left(D^4z^8+4D^4z^6+3D^8z^4+3z^4+4D^4z^2+D^4\right)}
\end{equation}
with the critical temperature for $L=3$ being
\begin{equation}\label{eq:CriticalTempL3}
     \frac{T_c}{A M}=\frac{1}{\ln z}\left(\frac{D^{4}\left(z^6-1\right)+z^2\left(z^2-1\right)}{D^4\left(z^6+1\right)+3z^2\left(z^2+1\right)}\right)^{2}
\end{equation}
The dependence of this critical temperature on coupling can be seen in Fig. \ref{fig:MF_ABMcomparison} and on number of layers in Fig. \ref{fig:criticalTemp_Layers}.
While this approach can be used any specific $L$, considering $L$-dimensional $\vec{x}$, the equations quickly become intractable with increasing $L$.\\

To make a general prediction for the critical temperature $T_c$ where the order in the system disappears, the states of all layers are assumed to be statistically the same.
This allows us to simplify our approach.
Instead of considering the exact microstate $\vec{x}_{ij}$ of the set of links between $i$ and $j$, we will consider a mesostate described only by the number of positive links $L^+$ among all $L$ links in the set.
Since by assumption all layers are statistically the same, the exact placement of positive and negative signs within the $L$ copies does not matter and the energy of a microstate depends only on the mesostate variable $L^+$, so dynamics of the system can be described entirely through that variable.
For a set of $L$ links, with $L^+$ positive and $L^-$ negative links, then $L^-=L-L^+$ and the mean polarization is,
\begin{equation}
    \langle x \rangle = \left\langle \frac{L^+ - L^-}{L} \right\rangle = \frac{2 \langle L^+\rangle}{L}-1
\end{equation}
where the $\langle x \rangle$ is the average polarization of links in each of the layers $\langle x \rangle = \langle x_{ij}^{(\alpha)} \rangle$, with mean over all links and ensemble, and $\alpha$ being any of the layers (they are statistically identical by assumption).
Because the number of microstates aggregated into a given mesostate $L^+$ is not fixed but is equal to the number of ways in which $L^+$ positive links are distributed across $L$ total, we need to consider this number of microstates contributing to a given mesostate $L^+$ when calculating canonical ensemble probabilities.
The number of microstates for mesostate $L^+$ is given by,
\begin{equation}
    \# microstates =  \binom{L}{L^+}
\end{equation}
The energy of state $L^+$ for a set of links between a given node pair is the sum of the Heider energy of links interacting with different layers as well as the Ising energy of interaction between different links in the set.
The Heider energy is according to Eq. (\ref{Eq:energy_raw_multi}), but instead of specific $x_{ij}^{(\alpha)}$ terms, we have $L^+$ positive and $L^-$ negative components in the sum over layers, and instead of all $x_{jk}^{(\alpha)}$ and $x_{ki}^{(\alpha)}$ terms we have mean-field polarization $x$
\begin{equation}
    E_{\mathrm{Heider}}(L^+)=-A(N-2)\left( L^+\langle x\rangle^2-(L-L^+)\langle x\rangle^2\right)
\end{equation}
One should be cautious not to mistake the macroscopic mean-field variable $\langle L^+ \rangle$ that appears in the expression for $\langle x \rangle$ with microscopic state variable $L^+$.
The distinction can be dropped only when calculating the mean value of microscopic $L^+$ later in Eq. (\ref{eq:meanLplus}).\\
The Ising energy corresponds to the sum over all pairs of links, $L^+$ of which are positive and $L^-$ negative, giving
\begin{multline}
    E_{\mathrm{Ising}}(L^+)=-K\frac{L^+(L^+-1)+(L-L^+)(L-L^+-1)}{2}\\
    +K L^+(L-L^+)
\end{multline}
with the complete energy being the sum of these two
\begin{equation}
    E(L^+)=E_{\mathrm{Heider}}(L^+)+E_{\mathrm{Ising}}(L^+)
\end{equation}
Hence the mean number of positive links can be calculated from the canonical ensemble, multiplying the factor for specific $L^+$ by the number of microstates such state actually represents.
At this point, the mean value of microscopic $L^+$ is considered the same as the mean-field variable $\langle L^+ \rangle$, which means we obtain a self-consistent equation for $\langle L^+ \rangle$.
\begin{equation}\label{eq:meanLplus}
    \left<L^+\right>=g(\left< L^+ \right>)= \frac{\sum_{L^+=0}^L L^+\binom{L}{L^+}e^{-E(L^+)/T}}{\sum_{L^+=0}^L \binom{L}{L^+}e^{-E(L^+)/T}}
\end{equation}
\begin{figure}
    \centering
    \includegraphics[width=\linewidth]{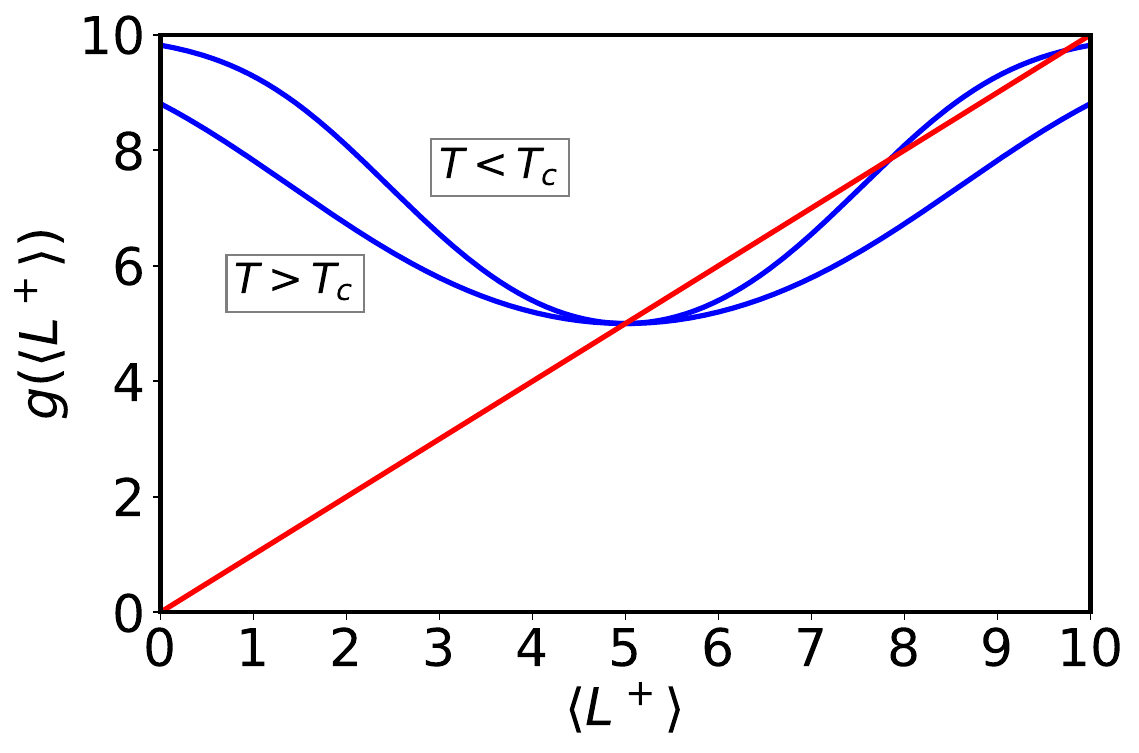} 
    \caption{ Graphical solution of the mean-field state equation (Eq. \ref{eq:meanLplus}) for $L=10$ layers. The solid blue curve is $y=g(\langle L^+ \rangle)$ and the red line is y=$\langle L^+\rangle$. When $T > T_c$, there is a single stable fixed point. When $T < T_c$, there are three fixed points, and the inner one is unstable}
    \label{fig:MF10}
\end{figure}
Here $\left\langle L^+\right\rangle=L$ is a paradise state, and $\left<L^+\right>=L/2$ is a completely disordered state with half the links positive and half negative.
Due to using a mesostate description, instead of $L$ equations, we have a simple scalar equation (Eq. \ref{eq:meanLplus}) that could be solved using a similar stability analysis approach.
Figure \ref{fig:MF10} shows a graphical representation of the right-hand side of Eq.(\ref{eq:meanLplus}), which solution is given by the intersection of the straight line $L^+$ and right-hand side.
For $T > T_c$, there is only one stable fixed point, $L^+ = L/2$, which corresponds to a disordered state of the system, while for $T < T_c$, an additional stable fixed point appears, corresponding to an ordered state, as well as an unstable fixed point separating basins of attraction for each stable solution. When $T = T_c$, both additional points are the same, unstable fixed point tangential to the diagonal line. The critical temperature can be numerically estimated from the fixed point equation  Eq.(\ref{eq:meanLplus}) and its derivative $d g(\langle L^+ \rangle)/d \langle L^+ \rangle=1$.\\
The $L$-coupled layers undergo a first-order transition at $T_c$ for any number of layers, which has been confirmed numerically for up to 10 layers using Eq.(\ref{eq:meanLplus}).

\subsection{Saturation of critical temperature}
The critical temperature $\frac{T_c}{AM}$ depends on $d$ (relative coupling strength) and it increases with an increase of this coupling strength, saturating for high $d$ values (see Fig.\ref{fig:MF_ABMcomparison}).\\
For large coupling strength $D \to +\infty$, Eq.\ref{eq:transcendentalEqn} and \ref{eq:CriticalTemp} can be reduced to
\begin{align}\label{11}
     8\ln z = \frac{(z^8-1)}{2z^4} \implies z \approx 1.72346\dots\\
      \frac{T_c}{A M}=\left(\frac{1}{\ln{z}}\right)\left( \frac{(z^4-1)}{(z^4+1)} \right)^2 \implies \frac{T_c}{AM} \approx 1.16516\dots
\end{align}
Comparing this to the results for a single layer, where $T_c/A M \approx 0.58258$ (in agreement with \cite{malarz_mean-field_2022}) we find that 
\begin{equation}\label{eq:bilayerCriticalTemp}
   T_c\big|_{L=2, d \rightarrow +\infty} \approx 2 \cdot T_c\big|_{L=1}
\end{equation}
To ensure the generality of the observed relationship (\ref{eq:bilayerCriticalTemp}) between the number of layers and the critical temperature, our study extends to a system with three layers, $L = 3$.
In the limit $D \to +\infty$, Eq.\ref{eq:transcendentalL3} and \ref{eq:CriticalTempL3} can be reduced to
\begin{align}
    8\ln z=\frac{\left(z^{12}-1\right)}{3z^{6}} \implies z \approx 1.43749\dots\\
     \frac{T_c}{A M}=\left(\frac{1}{\ln{z}}\right)\left( \frac{(z^6-1)}{(z^6+1)} \right)^2 \implies \frac{T_c}{AM}\approx 1.74516\dots
\end{align}
The obtained saturation critical temperature satisfies the relation
\begin{equation}\label{eq:3layerCriticalTemp}
   T_c\big|_{\alpha=3, d \rightarrow +\infty} \approx 3 \cdot T_c\big|_{\alpha=1}
\end{equation}
For $L$ coupled layers, the transcendental equation at $D \to +\infty$ could be used to obtain the saturation critical temperature.
\begin{align}
    8\ln z = \frac{\left(z^{4L}-1\right)}{L z^{2L}}\\
     \frac{T_c}{A M} = \left(\frac{1}{\ln{z}}\right)\left( \frac{(z^{2L}-1)}{(z^{2L}+1)} \right)
\end{align}
This equation allows us to conclude that
\begin{equation}\label{eq:CritTemp_Layernumber}
    T_c\big|_{\alpha=L, K\rightarrow +\infty} \approx L \cdot T_c\big|_{\alpha=1}
\end{equation}
Thus the saturation critical temperature increases proportionally with the number of layers $L$

\section{Numerical Simulations of paradise state stability}
\begin{figure}
    \centering
    \includegraphics[width=\linewidth]{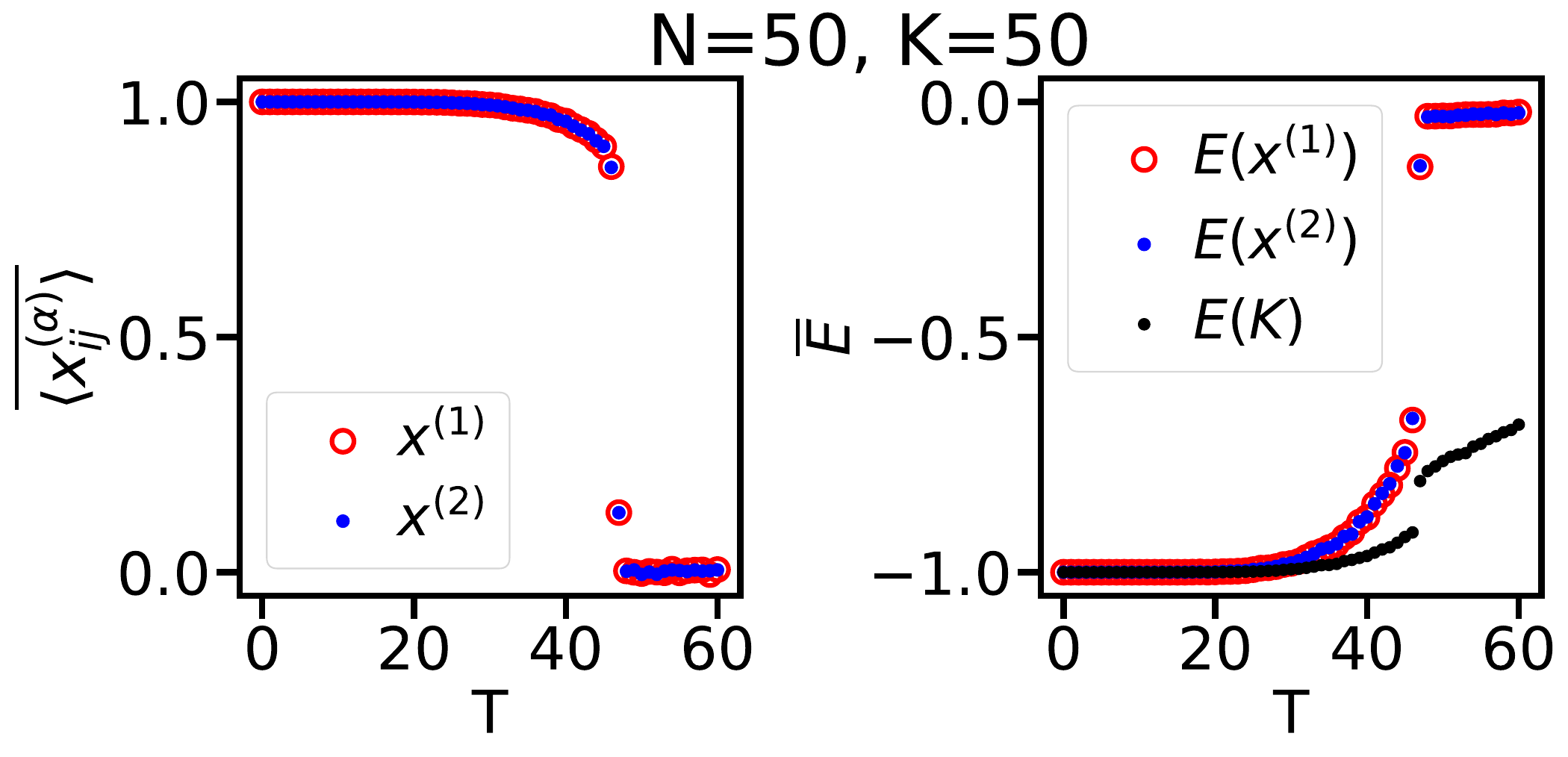} 
    \caption{Mean link polarization $x^{(1)}= \langle x_{ij}^{(1)} \rangle$ and  $x^{(2)}= \langle x_{ij}^{(2)} \rangle$ for a duplex network as a function of temperature T for the complete graphs with $N = 50$ nodes and a coupling strength $K=50$ (corresponding to d = 1). The points show averages over  50 independent simulation. Initial conditions were paradise states at both layers, where the red and blue symbols correspond to layer one and two respectively. The corresponding energy is shown in the right panel. In contrast to the intralayer energies (red and blue), the interlayer energy (black) increases slowly when $T>T_c$, with layers remaining partially synchronized even when intralayer disorder sets in.}
    \label{fig:Pol_En_PP}
\end{figure}

\begin{figure}
    \centering
    \includegraphics[width=\linewidth]{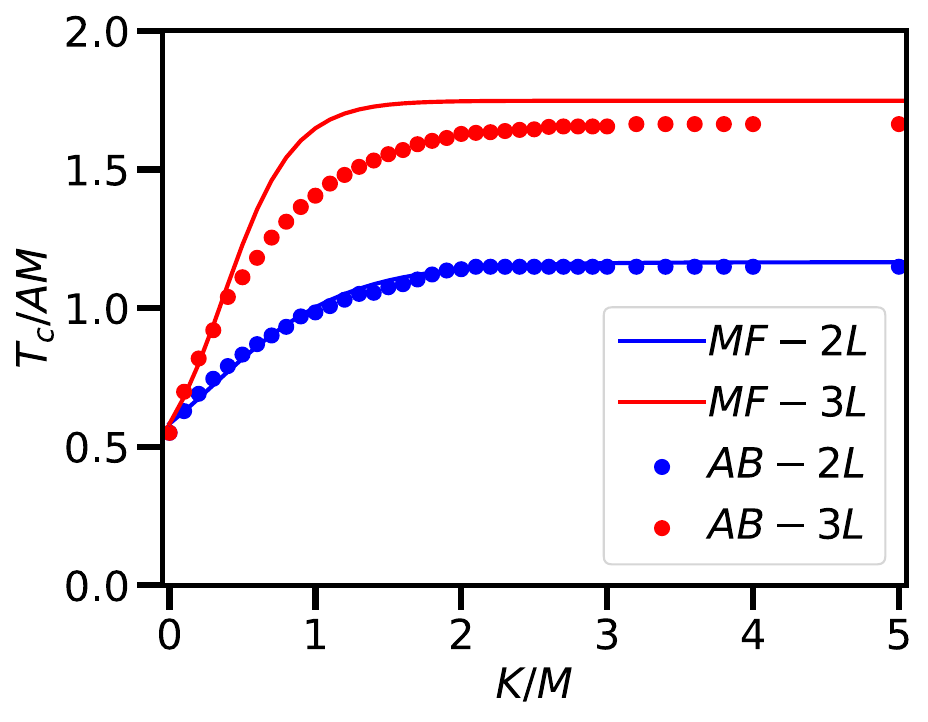} 
    \caption{Critical temperature $T_c$ increases with coupling strength $K$ and saturates to $T \approx L\cdot T_{c,L=1}$ for high $K/M$ values. Lines show mean-field solution (Eqs. \ref{eq:CriticalTemp},\ref{eq:CriticalTempL3}), and points show results of numerical Monte-Carlo simulations.}
    \label{fig:MF_ABMcomparison}
\end{figure}

\begin{figure}
    \centering
    \includegraphics[width=\linewidth]{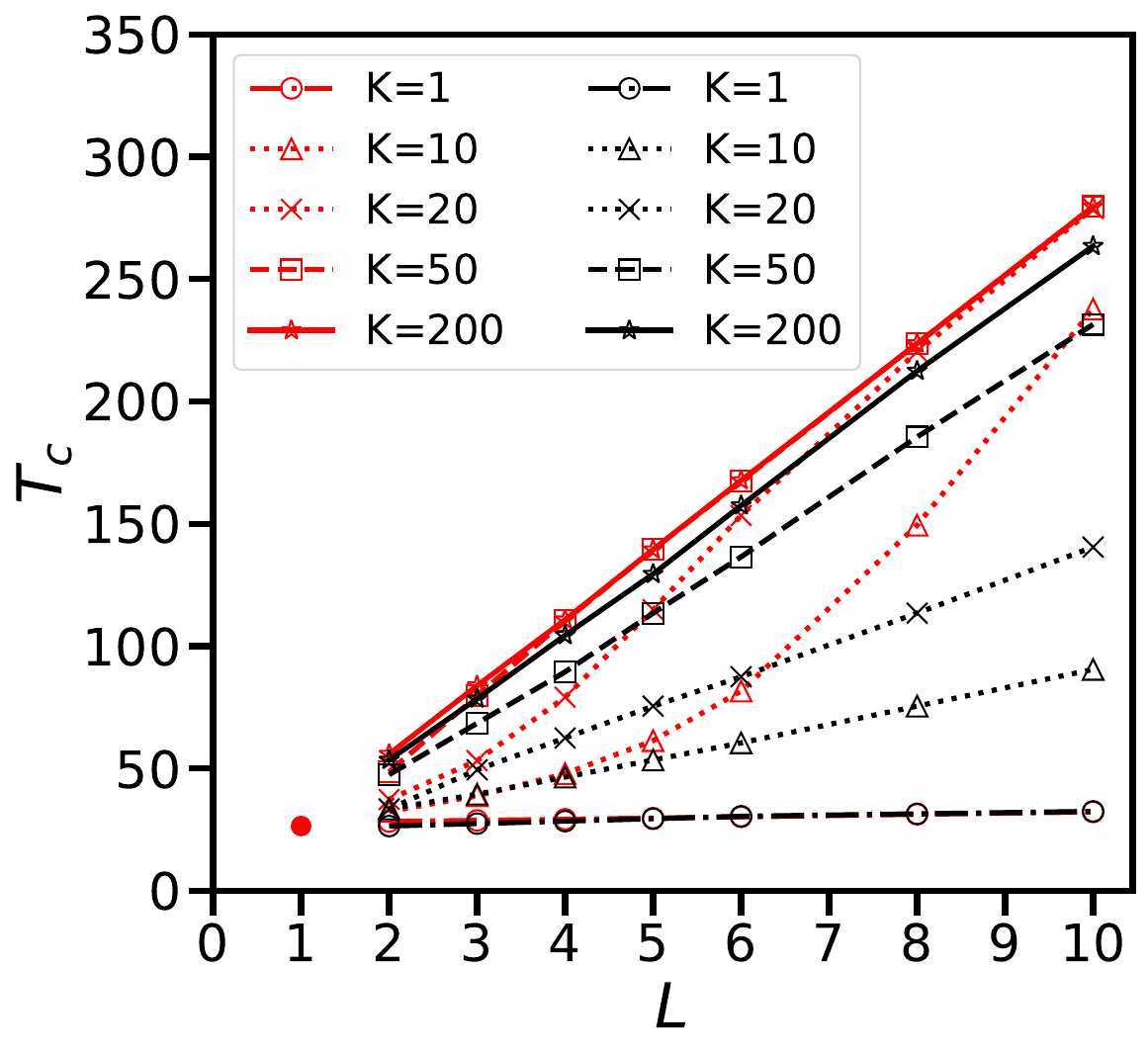} 
    \caption{The critical temperature $T_c$ increases with the number of layers $L$ when coupling $K$ is positive. Black symbols and lines show results of Monte-Carlo simulations, while red symbols and lines show mean-field prediction (iterating Eq.\ref{eq:meanLplus} as a map) at different interlayer coupling strengths for $N=50$ and $A=1$. The critical temperature for a single layer is shown as a red symbol at $T_c=26.5$.}
    \label{fig:criticalTemp_Layers}
\end{figure}

For the verification of  the analytical calculations presented in the previous section, numerical Monte Carlo simulations of the model have been performed.
The Metropolis algorithm is used with single random link sign change as elementary updates and asynchronous update with $L N (N-1)/2$ elementary updates as one time step.
The initial condition is always started with a fully connected graph in the \emph{paradise} configuration (all links positive).
In the case of $L>2$, a different setup has been employed, where an entire set of links described by $\vec{x}_{ij}$ is updated using a heat-bath method as an elementary update, with $N(N-1)/2$ updates constituting one time step.
This is because for high coupling $K$, any changes to a single sign are very unlikely in the single-link update Metropolis method.
Effectively, there is a large energy barrier to switching the link set to the opposite as a whole, which causes simulations to become extremely slow for $L \geq 4$.
The simulation results for this approach are the same as for single-link Metropolis update for $L=2$.\\
For the complete graph, the number of pair neighbors of nodes $ij$ is equal to $M=N-2$ and we keep $A^{(1)}=A^{(2)}=1$.
The intralayer energy has been rescaled to the average value of Heider energy per triad whereas the interlayer energy is scaled to the average energy value per interacting link pair. \\
The simulations are repeated for a range of temperatures from $T=0$ with step $\Delta T$ to find the highest value of $T$ for which $\langle x_{ij}^{(\alpha)}\rangle$ is positive. The true value of $T_c$ lies in the interval $[T^*, T^* + \Delta T]$, and the estimated value of critical temperature is $T_c = T^*+\Delta T/2$.\\
For the initial condition of paradise state, the system moves to a disordered state above temperature $T_c$.
Fig.\ref{fig:Pol_En_PP} shows that both layers behave in the same, aligned way.
The intralayer energies $E(x^{(1)}), E(x^{(2)})$ approach zero above the critical temperature, and the mean polarization values fluctuate around zero.
The interlayer energy $E(K)$ is negative both above and below the critical temperature, showing that the link signs in different layers are partially aligned even in disordered state above $T_c$.\\
The critical temperature $T_c$ predicted from numerically evaluated analytical mean-field theory is compared to those obtained from Monte-Carlo simulations (Fig. \ref{fig:MF_ABMcomparison}). In spite of the slight shift of the transition when comparing numerical simulations with that of mean-field approximation, the analytical method is found to be  successful.
Fig.\ref{fig:criticalTemp_Layers} shows how the critical temperature varies with the number of layers where the simulations also validate Eq.\ref{eq:CritTemp_Layernumber} in the saturation limit.
Here, there is some discrepancy between our predictions and simulation results for a large number of layers.
The analytical approach predicts that as $L$ increases, even if coupling $K$ is small, at some point the critical temperature will start rising non-linearly, increasing all the way up to the same saturated critical temperature $T_c$, regardless of $K$.
Even low $K$ values show this behavior for a sufficiently large number of layers $L$.
This behavior can be best seen in Fig. \ref{fig:criticalTemp_Layers} for $K=10$ and $K=20$.
Simulations show, however, that no such transition exists, and the critical temperature increases linearly with $L$ for any $K$, with a slope depending on $K$.
This means that our mean-field approach is unable to fully and accurately predict the behavior of the system with many weakly interacting layers, although it is still qualitatively correct.
Our predictions work both for low $K$ and $L$, where the analytical approach predicts a linear increase of $T_c$ with $L$, as well as high $K$ values, where the temperature is essentially always saturated.

\section{Critical behavior of the system with layers in different states}\label{sec:mismatched}
The critical temperature investigated so far was defined as the temperature where the non-zero stable state disappears.
The analytical approaches only describe the existence of a paradise state and look into its stability and transition into disorder.
The multistability of the system means, that in the range of parameters where paradise was stable, other states may exist, and starting from different initial conditions, the system may settle into different stable states.
In this section, we explore what happens in a duplex network in more detail, focusing on situations where the two layers are in different configurations.\\
For a single layer, the three prominent possible states are a paradise, a disordered state, and what we will call a \emph{two-clique state}, where vertices are split into two groups with all internal links in these groups positive, and all links between groups negative.
The entire duplex network can also exist in three different states: matching order, mismatched order, and disorder.
What we call the \emph{matching order} state encompasses all states where link signs in each layer are balanced according to Heider balance dynamics, such as paradise or two-clique state, and the ordering of layers matches -- both layers are in paradise state, or the split into groups in both layers are the same.
Note that the matching order state is broad and encompasses both paradise and matching two-clique states since they behave in the same way.
The \emph{mismatched order} state means that each layer is internally ordered, but the ordering does not match -- for example, one layer is in a paradise state while the other is in a two-clique state.\\
It is worth noting that the division into two cliques is the most likely outcome of balance dynamics starting from a random state \cite{antal2006social} provided the thermal noise is not enough to keep it disordered.
The analytical approaches introduced before, however, do not take its existence into account at all.
The conclusions from an in-depth numerical investigation presented here may differ significantly from mean-field predictions whenever two-clique states are involved.
The most important difference is that the mean field predicts a disordered state to be always stable, even at low temperatures, which is not true in reality, as is shown later in this section.
In simulations, the system orders into a two-clique state under such conditions, that also possess zero polarization, just like the disordered state.\\

We consider and investigate numerically two scenarios where the initial condition is mismatched order state:
\begin{itemize}
    \item One layer is in the paradise state, while the other is a two-clique balanced state
    \item Both layers are in a two-clique balanced state, but the cliques don't match between layers
\end{itemize}
Based on the results obtained in these situations as well for initially disordered system, we draw conclusions regarding what system states exist in what range of parameters, and what are transitions between them.

\subsection{Coupling between a paradise and two-clique state}
\label{sec:p2c}
\begin{figure}
    \centering
    \includegraphics[width=\linewidth]{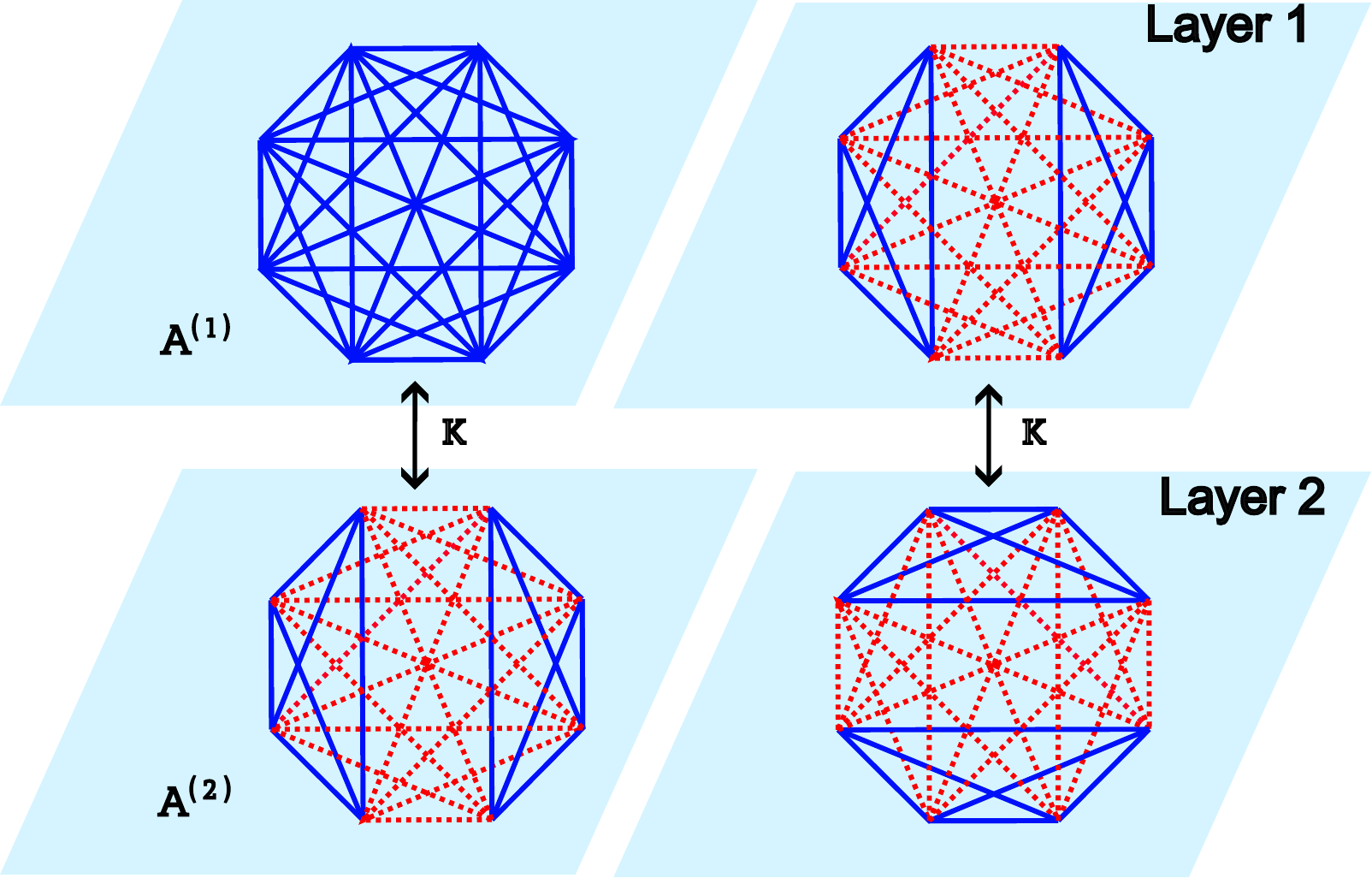} 
    \caption{An example of coupling between two layers in a system in a mismatched order state. The left part shows a paradise state in one layer and a two-clique state in the second layer (Sec. \ref{sec:p2c}), while the right part shows layers in orthogonal two-clique states (Sec. \ref{sec:2c2c}). The same nodes exist in both layers, and interlayer links connect only the same link replicas with coupling strength $K$. Blue links represent positive links and dotted red lines represent negative links.}
    \label{fig:P2C_illustration}
\end{figure}

\begin{figure}
    \centering
    \includegraphics[width=0.8\linewidth]{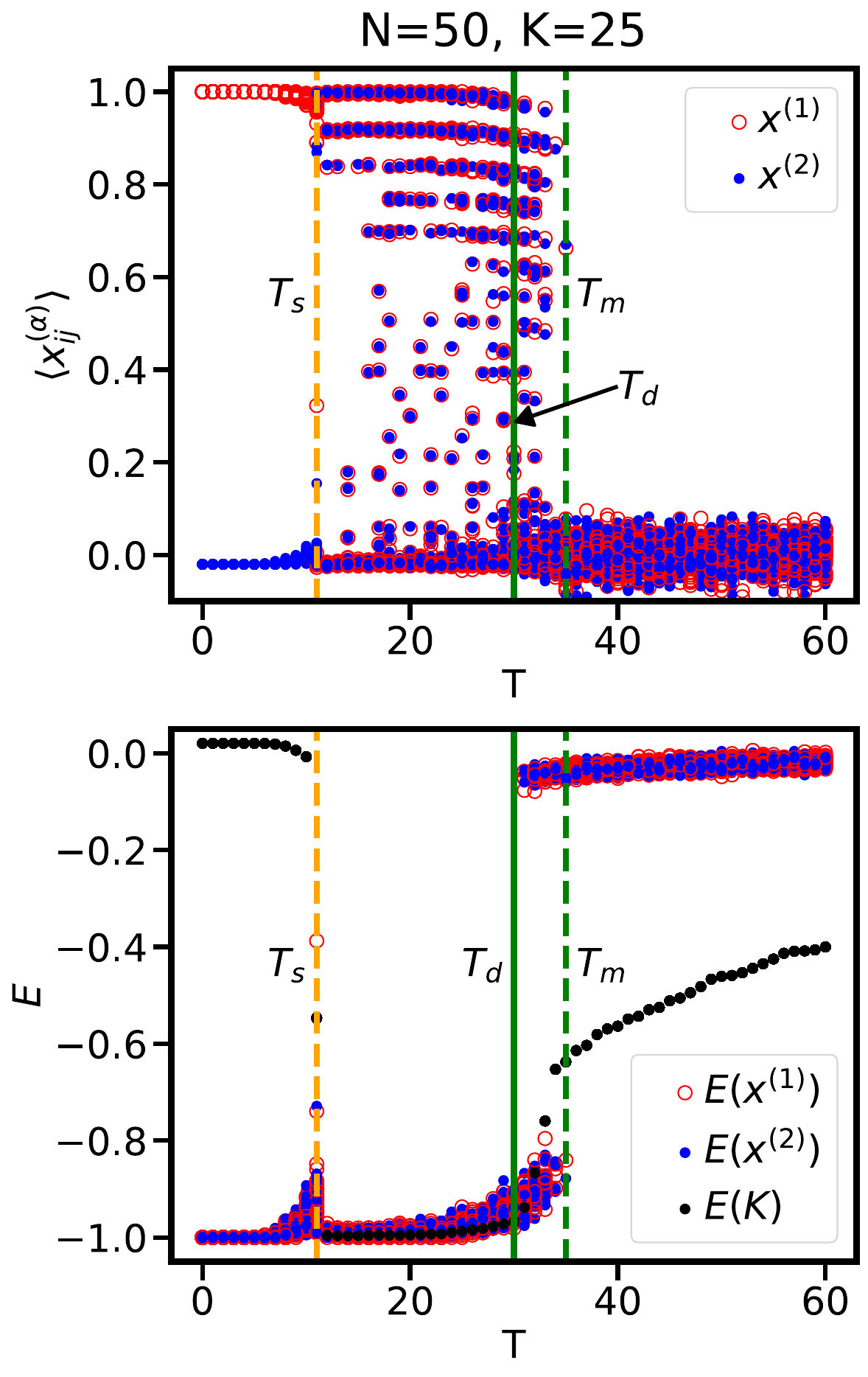} 
    \includegraphics[width=0.8\linewidth]{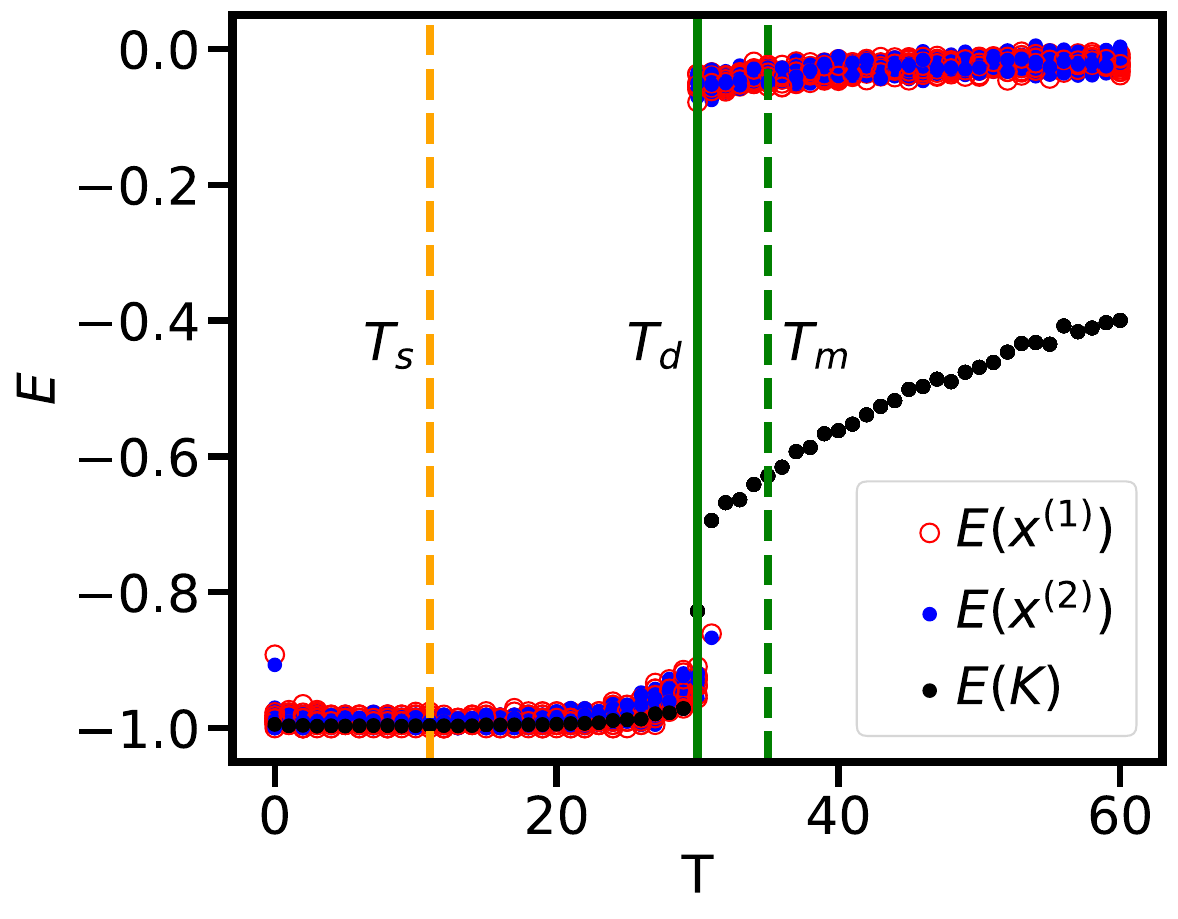} 
    \caption{In the case of coupled paradise and two-clique states, the system shows two transitions: synchronization of layers starting at $T_s$ and transition to disorder at $T_d$, with some instances retaining synchronized order up to $T_m$. The top graph shows the mean link polarization of $50$ independent simulations over a range of temperatures, with one layer starting in the paradise state (red circles) and the second layer starting in the two-clique state (blue points). The middle graph shows the corresponding mean intralayer energy (red and blue) and interlayer energy (black). The bottom graph shows intralayer and interlayer energy for a duplex network starting in a disordered state, showing that spontaneous order appears at any temperature below $T_d$.}
    \label{fig:P2CPol_En}
\end{figure}

Consider a mismatched order state containing paradise in one layer and a two-clique state in the second, as shown in Fig.\ref{fig:P2C_illustration}.
Such a system possesses layers that are internally in the ground state at low temperatures (this is not a ground state of the entire system due to interlayer energy), with average polarization $\langle x_{ij}^{(\alpha)}\rangle$, fluctuating around $1$ for the paradise layer and $0$ in the two-clique layer.\\ 
Assume that the size of the entire group is an even number $N = 2m$ and that each hostile group has the size $m$. Each node possesses $\left( m-1 \right)$  positive links to agents in its own clique and $m$ negative links to the second clique. 
We provide evidence that in such a scenario, the system of two layers undergoes phase transitions at two different temperatures $(T_s, T_d)$, with additional characteristic temperature $T_m$, as shown in Fig.\ref{fig:P2CPol_En}.

\begin{figure}
    \centering
    \includegraphics[width=\linewidth]{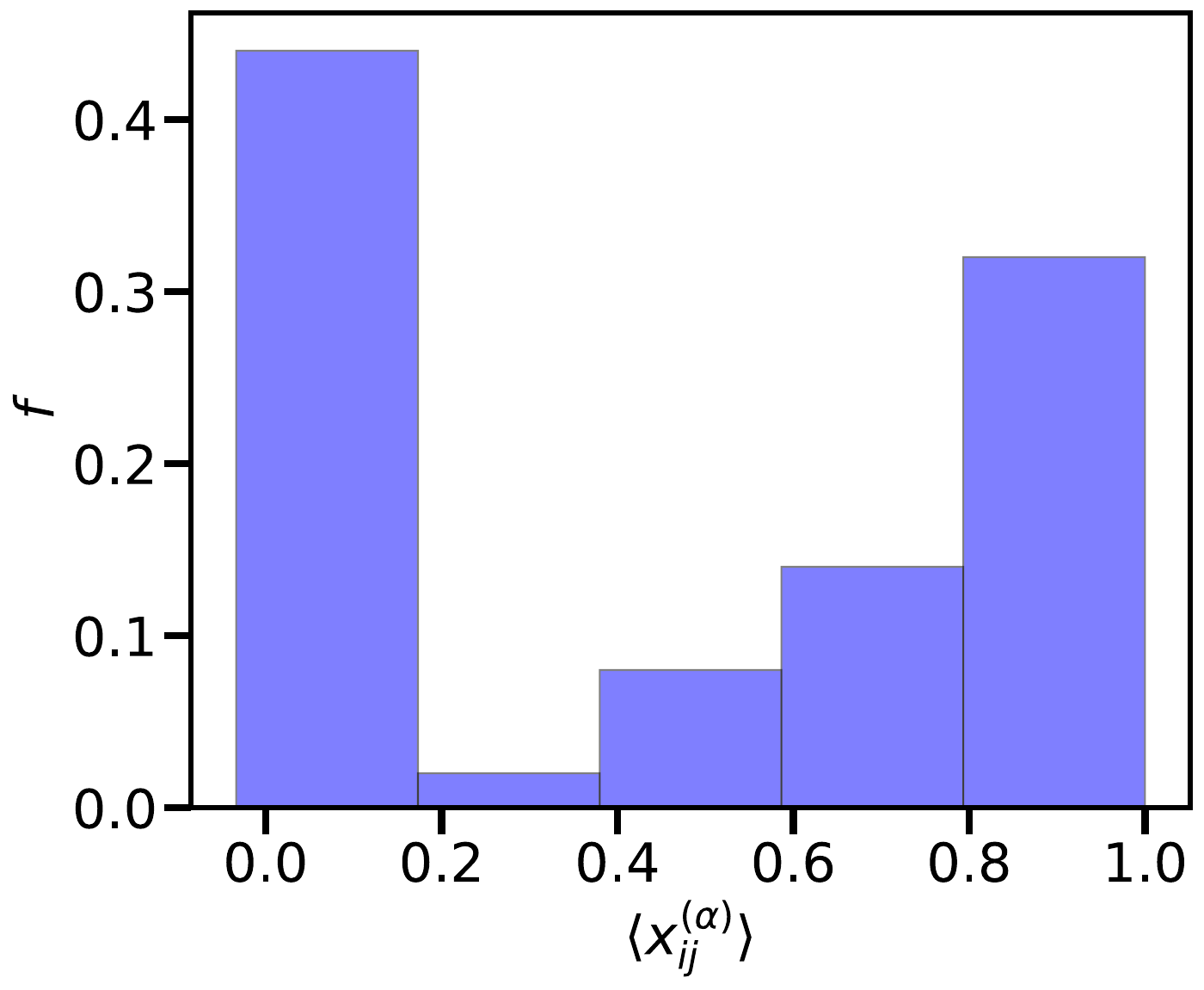} 
    \caption{When paradise and two-clique states synchronize, it results most often in either paradise $\langle x_{ij}^{(\alpha)}\rangle \approx 1$ or two-clique state with cliques of equal sizes $\langle x_{ij}^{(\alpha)} \rangle \approx 0$, but sometimes different size cliques also appear. The graph shows the distribution of mean links polarization after paradise and two-clique states synchronize at the temperature $T$ where $T_s < T < T_d$.}
    \label{fig:freqP2C}
\end{figure}

The temperature $T_s$ is the temperature above which the mismatched order loses stability, pushing the layers to synchronize with each other and create a matching order state.
Above $T_d$ the disordered state becomes stable, so that it becomes an option for the final state of evolution of the initial mismatched order.
In this situation, whether the system ends up in a matching order state or a disordered state is essentially random.
As the temperature increases further, the disordered state is a more and more common outcome, up to a characteristic temperature $T_m$, above which disorder is always the final outcome of the initial mismatched order state.
Note that $T_m<T_c$, so the initial mismatched order evolves to disorder at a lower temperature than the initial matching order, which can persist up to temperature $T_c$.
We consider $T_s$ and $T_d$ as critical temperatures, since certain states gain or lose stability, while at $T_m$ no such thing happens.
Hence we don't refer to $T_m$ as a critical temperature but as a characteristic temperature.\\
The outcome of synchronization of the initially mismatched order state is not always the same and contains randomness.
In our case of a mismatched paradise and two-clique state, the outcome could be either a matching paradise or a matching two-clique state.
In addition, the resulting two-clique state does not need to match the state of the initial two-clique layer -- the cliques could be of different sizes and placement than initially, which suggests both layers may undergo reconfiguration, instead of one layer imposing its own state on the other.
The mean link polarization of a two-clique state depends on the sizes of cliques $m_1$ and $m_2$
\begin{equation}
    \langle x_{ij}\rangle= \frac{\binom{m_1}{2} + \binom{m_2}{2} - m_1 \cdot m_2}{\binom{N}{2}}
    \label{eq:2cliquePol}
\end{equation}
$\langle x_{ij}\rangle$ would be near $1$ for clique sizes $m_1 \gg m_2$, and it would be close to zero for $m_1 \approx m_2$.
We quantitatively investigated the frequency $f$ of mean link polarization for both network layers after the synchronization event.
Fig.\ref{fig:freqP2C} shows the distribution of mean link polarization. 

Even though the paradise state $\langle x_{ij}\rangle=1$ or two-clique state with equal clique sizes $(\langle x_{ij}\rangle=0)$ are most likely results of the synchronization, two cliques of different sizes are also a probable result.

\begin{figure}[t]
    \centering
    \includegraphics[width=\linewidth]{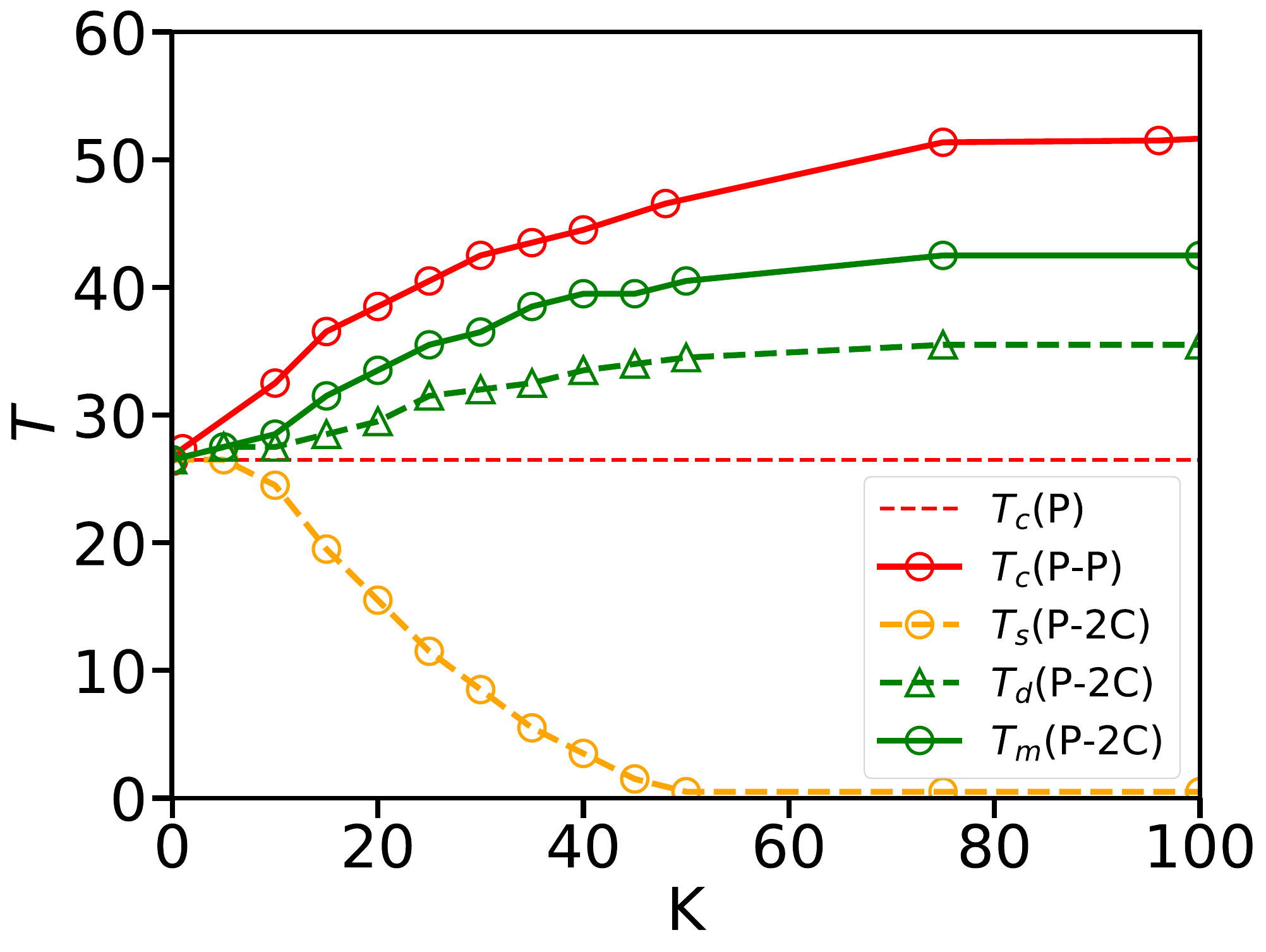} 
    \caption{The synchronization temperature $T_s$ (orange), the limit of spontaneous order $T_m$ (dashed green), the limit of retaining order after synchronization $T_d$ (green), and critical temperature for the existence of ordered states $T_c$ as a function of the Ising interaction strength between layers. Data for agent-based simulations for $N=50$ and $A=B=1$ with $50$ realizations. Here $P$ and $2C$ represent paradise and two-clique initial conditions respectively. All temperatures saturate around $K\approx 80$ for these parameters.}
    \label{fig:criticaltemperatures}
\end{figure}

\begin{figure*}[htb]
    \subfloat[\label{subfig:a}]{
  \includegraphics[width=0.3\textwidth]{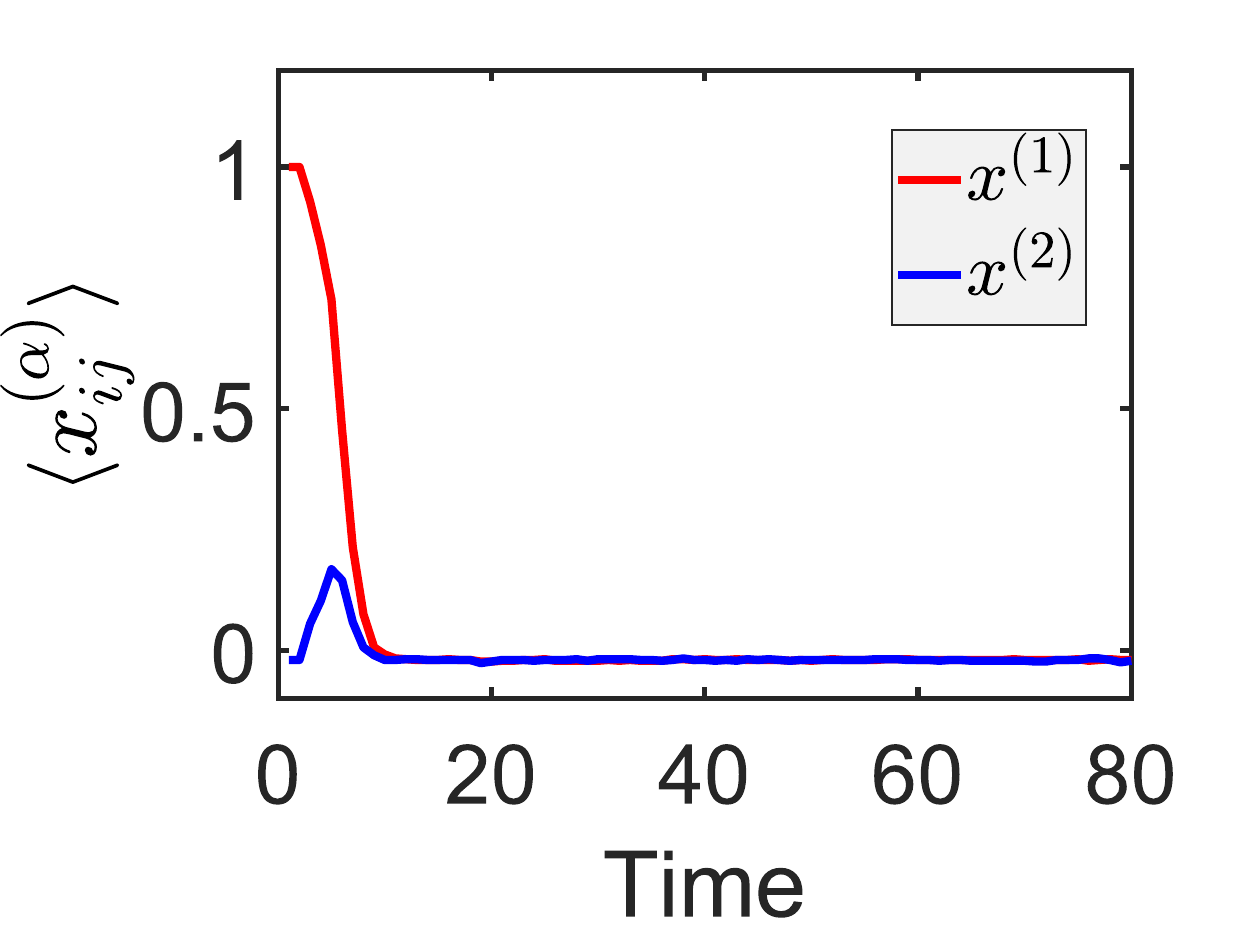} 
    }\hfill
    \subfloat[\label{subfig:b}]{
  \includegraphics[width=0.3\textwidth]{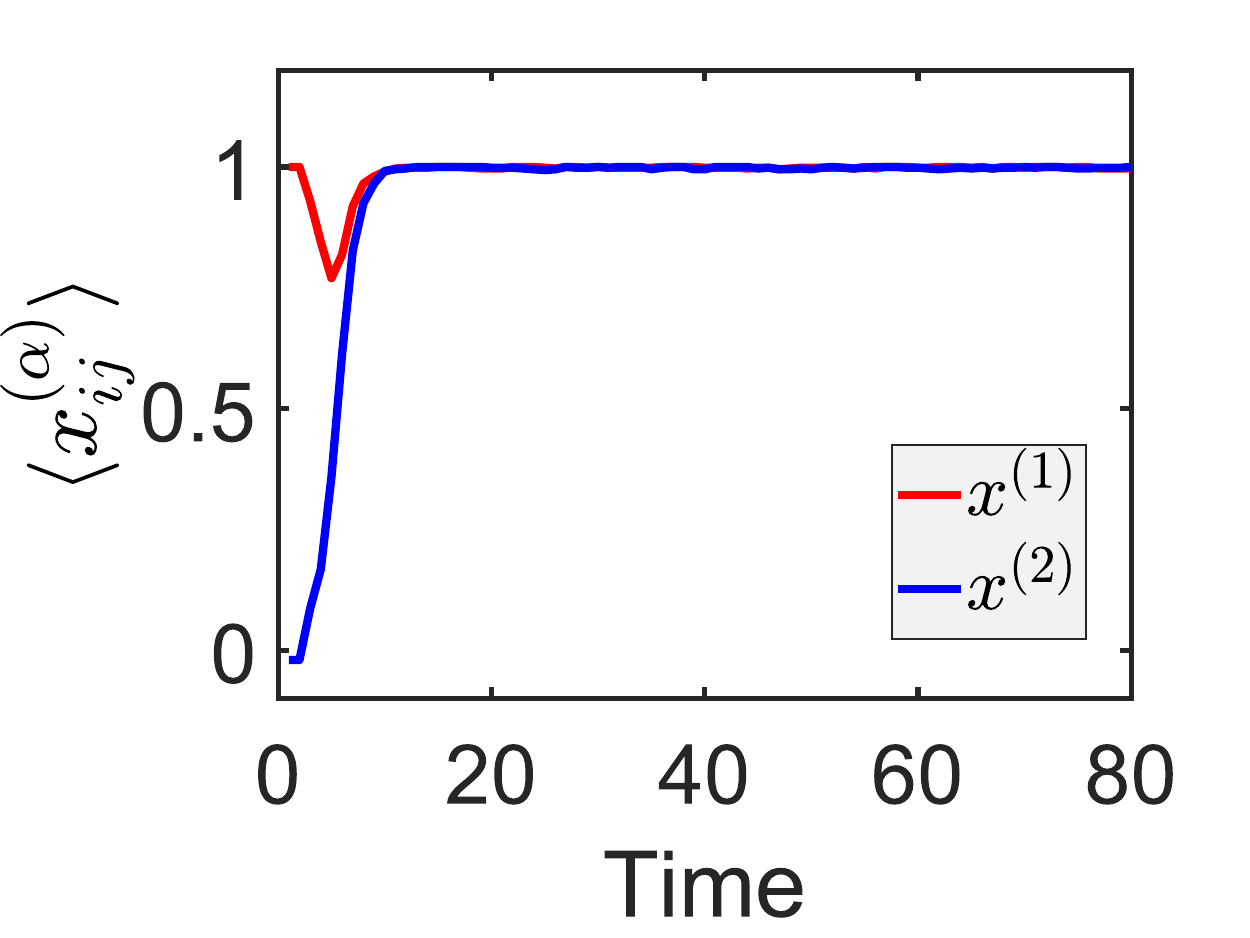} 
    }\hfill
    \subfloat[\label{subfig:c}]{
  \includegraphics[width=0.3\textwidth]{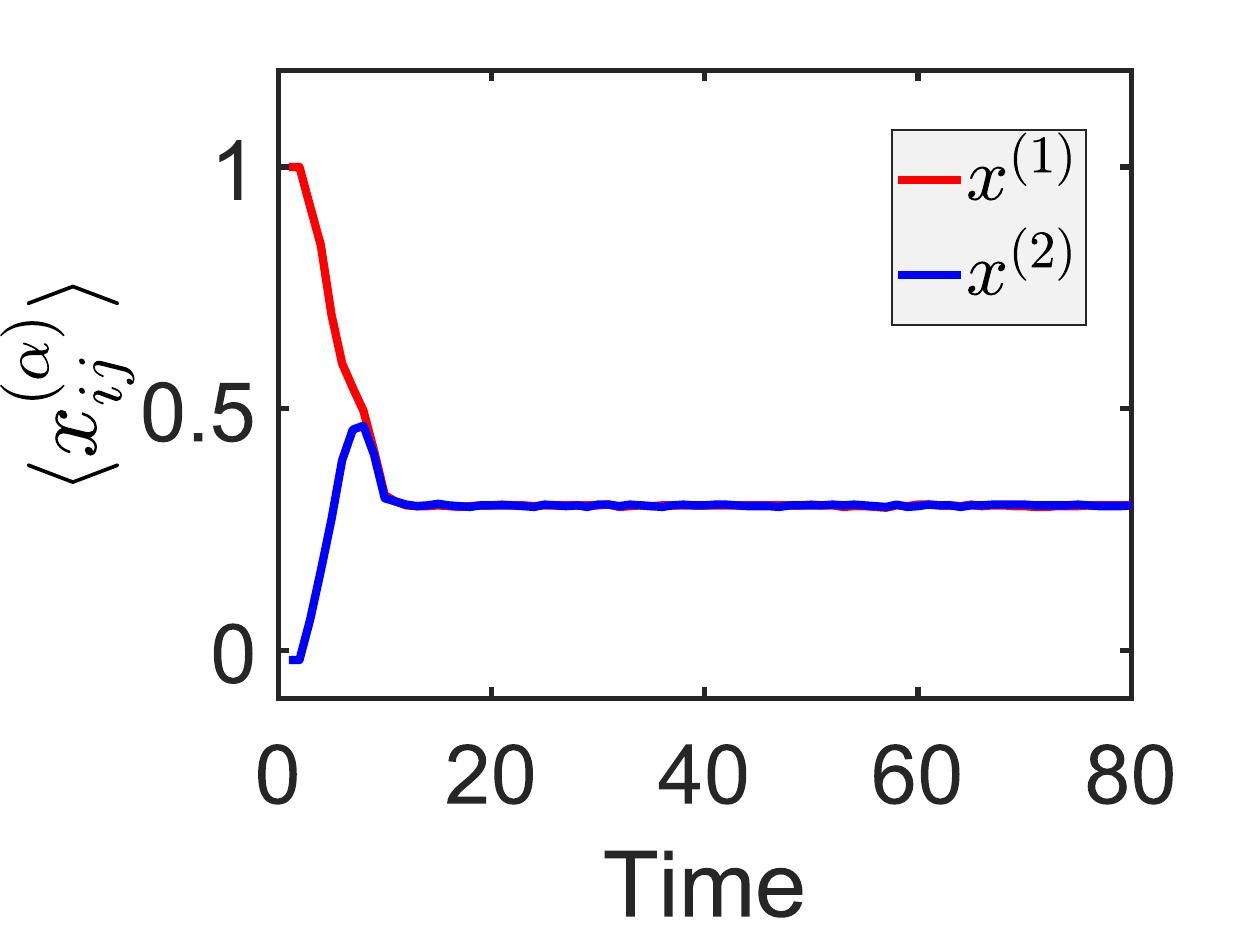} 
    }
    \caption{Final state of the synchronization between two mismatched layers is partially random due to thermal fluctuations. Example trajectories for the synchronization scenario between paradise (red) and two-clique (blue) layers for $N$=50, $K$ = 25 and $T$=20. The synchronized state can be either a two-clique state similar to the initial condition of the second layer (panel (a)), a paradise state (panel (b)), or a two-clique state of different sizes, matching neither of the two initial layer states (panel (c)). Each shown scenario is a result of independent agent-based simulation under the same initial conditions and parameters.}
    \label{fig:PolznTraj}
\end{figure*}

Fig. \ref{fig:criticaltemperatures} shows how temperatures $T_s$, $T_d$ and $T_m$ depend on coupling strength between layers, with the initial state being a mismatched state composed of paradise and two-clique states ($P-2C$).
These temperatures are compared to the critical temperature $T_c$ determined from simulations with paradise--paradise ($P-P$) initial state (which is the same for any matching order state).
Increasing the coupling strength decreases the critical temperature for synchronization while the critical temperature for disorder increases and saturates for large coupling strength.
Note that when the system results in a disordered state, it is still partially synchronized due to the interlayer Ising interactions.
This is equivalent to paramagnetic partial ordering due to the external field, which in this case is the impact of the other layer.
In effect, the interlayer energy $E(K)$ in Fig. \ref{fig:P2CPol_En} only asymptotically approaches zero with increasing temperature, unlike intralayer energy related to internal layer ordering.\\
An interesting observation during synchronization is how the link polarization of the two layers evolves over time.
Fig.\ref{fig:PolznTraj} displays examples of time trajectories of mean link polarization during relaxation processes.
The two layers, initially in paradise state and two-clique state with same-sized cliques, synchronize to one of the three possible matching order states: paradise, two-clique state with same-sized cliques, or two-clique state with cliques of different sizes (see also Fig.\ref{fig:PolznTraj}).

\subsection{Coupling between different two-clique states}
\label{sec:2c2c}
\begin{figure}
    \centering
    \includegraphics[width=\linewidth]{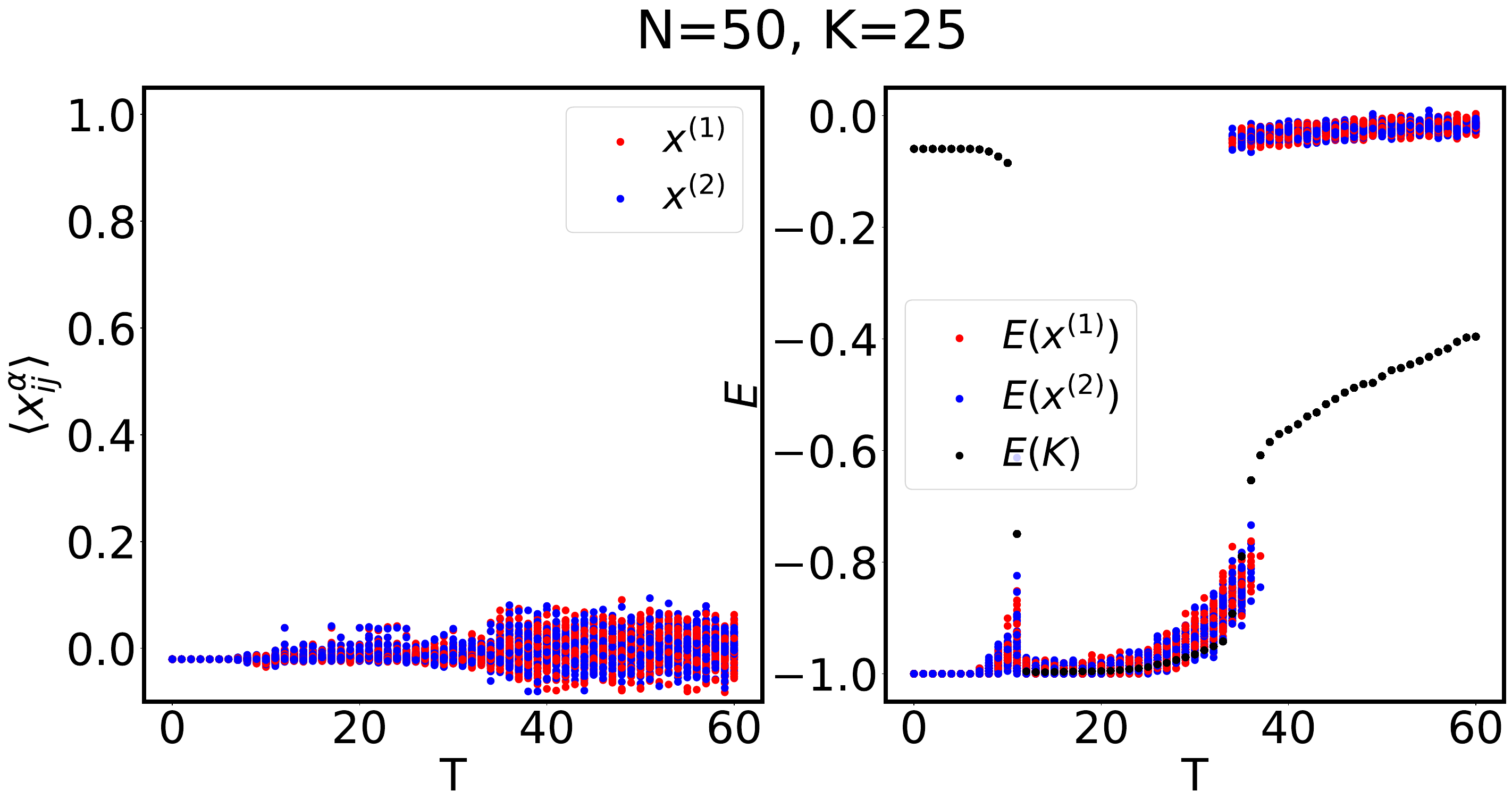} 
    \caption{Behavior of mismatched order of two orthogonal two-clique layers closely resembles mismatched order of paradise and two-clique layers seen in Fig. \ref{fig:P2CPol_En}. The left panel shows the variation of mean link polarization for two orthogonal two-clique layers over temperature for 50 independent simulations. The right panel shows the corresponding variation of mean intralayer (red and blue) and interlayer (black) energy over temperature.}
    \label{fig:2C2C}
\end{figure}

Another possibility for mismatched order is to have two-clique states in both layers, but with the two-clique states being "orthogonal" to each other.
The cliques are said to be orthogonal if they are maximally non-matching, as seen on the right in Fig. \ref{fig:P2C_illustration}.
Every clique in the first layer is evenly split into two opposing cliques in the second layer, resulting in half of all node pairs having opposite link signs in different layers.\\
If the cliques match, then the behavior is very similar to the paradise-paradise initial condition, except the mean link polarization is not $1$, but depends on the relative sizes of the cliques, down to $0$ for cliques of the same size.
The critical temperature of transition to disorder is also the same $T_c$ as for the matching paradise state.\\
For the mismatched order of two-clique states in each layer, the behavior closely  resembles that of the mismatched order of paradise and two-clique situation described in the previous subsection.
There are two critical temperatures: synchronization at $T_s$ and appearance of disorder at $T_d$, as well as the same characteristic temperature $T_m$ above which mismatched order stops synchronizing and always falls into disorder, as shown in Fig. \ref{fig:2C2C}.

\subsection{Stable states and transitions}
Evidence gathered and shown in previous subsections allows us to paint an overall picture of possible system states, when they are stable and when they aren't, as well as what transitions between them can occur.
\begin{figure}
    \centering
    \includegraphics[width=0.8\linewidth]{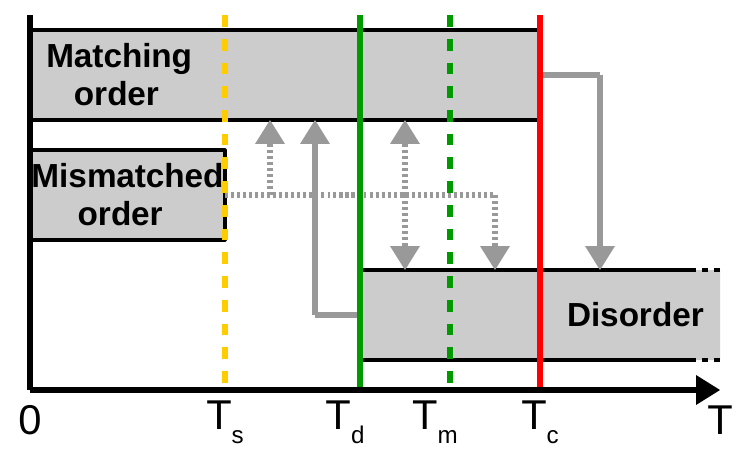} 
    \caption{The Heider Balance dynamics on a duplex network is multistable, with different states possible depending on the temperature. Grey boxes show a range of temperature where a given state is stable, with grey arrows showing possible transitions between these states. Four characteristic temperatures can be distinguished: synchronization at $T_s$, the appearance of disorder at $T_d$, the limit of mismatched order state becoming matching $T_m$, and the disappearance of order at $T_c$. Note that ordered states include both paradise and two-clique states and the transition to order almost always results in a two-clique state.}
    \label{fig:phasediagram}
\end{figure}
Fig. \ref{fig:phasediagram} shows schematically what states are stable in what temperatures.
At low temperatures, both mismatched and matching order states are stable, with intralayer ordering including either a two-clique state or paradise.
Temperature $T_s$ is the upper bound for the stability of mismatched order, temperature $T_d$ is the lower bound for the stability of disorder, and temperature $T_c$ is the upper bound for the stability of the matching order state.
If the temperature of an already existing matching order state increases above $T_c$, the system transits into a disorder.
Similarly, if the temperature for an existing disordered state is lowered below $T_d$, it transits into a matching order state.
It's worth noting that the probability that a self-organized matching order state will be a paradise is vanishingly small for larger systems.
If the temperature of a system in a mismatched order state is raised above $T_s$, the transition that happens depends on the new temperature.
For $T \in (T_s,T_d)$, it will always become a matching order -- the only state stable under these conditions.
For $T \in (T_d,T_m)$, it may either become a matching order or disorder, as both are stable.
For $T > T_m$, the result will always be a disorder, despite matching order being a stable state up till $T_c > T_m$.
This means for $T \in (T_m,T_c)$, while it is possible for an already existing matching order to persist, thermal fluctuations are too large for such a state to appear. 
Note that this summary only covers a duplex network with $A^{(1)}=A^{(2)}$.
If Heider interaction strengths of layers can be different, the situation becomes much more complex.

\section{Conclusions}
In this paper, we explore phase transitions of a multistable thermalized agent system on a multiplex network, with Heider balance intralayer dynamics and Ising interlayer coupling.
The system displays several interesting transitions (see Fig. \ref{fig:phasediagram}).\\
The possible system states include:
\begin{enumerate}
    \item Matching order state, where each layer has the same low-energy ordered state -- two-clique state or paradise
    \item Mismatched order state, where each layer is internally ordered, but the configuration in each layer is different, resulting in higher interlayer interaction energy. An example of such a state would be a paradise state in one layer and a two-clique state in the second
    \item Disordered state, dominated by thermal noise
\end{enumerate}
We identified three critical temperatures (see Fig. \ref{fig:phasediagram}):
\begin{enumerate}
    \item Critical temperature $T_c$ above which no ordered states are stable. This temperature is applicable for any number of layers.
    The mean-field estimation for $T_c$ is in good agreement with the agent-based simulation results for various number of layers and interlayer coupling strengths.
    For a system of $L$ layers, the temperature $T_c$ saturates with the strength of positive interlayer coupling, up to $(T_c)_L=L\cdot (T_c)_{L=1}$.
    While the mean-field approach only considers fluctuations around the paradise state, numerical results indicate that a state of matching two-clique layers also loses stability above the same $T_c$.
    \item Critical temperature $T_s$ for the existence of mismatched order states. This temperature is well-defined only for duplex networks and would split into multiple different temperatures for $L>2$.
    Mismatched order states are only stable below temperature $T_s<T_c$.
    Above $T_s$ thermal fluctuations allow the system to relax from a higher energy mismatched order state to a matching order state with a lower energy, similar to freezing by heating phenomenon \cite{helbing2000freezing, holyst2023history}.
    If the temperature is high enough, thermal noise may force the system into a high entropy, disordered state instead.
    \item Critical temperature $T_d$ below which the system orders spontaneously. This temperature is applicable for any number of layers, although it has been fully investigated for duplex networks only.
    A disordered system at $T<T_d$ spontaneously orders, forming a two-clique matching order state.
    There is a vanishingly small probability for a larger system to order into a mismatched order state (if $T<T_s$) or for the matching order state to be a paradise, that can be considered a special case of two-clique state with one clique being of size zero.
\end{enumerate}
In addition, there is a characteristic temperature $T_m$, which is an upper boundary for mismatched order state to synchronize into matching order state.
Above $T_m$ the transition from mismatched order will always be towards a disordered state.

Note that all the critical temperatures $T_s$,$T_d$,$T_c$, as well as the characteristic temperature $T_m$ depend on other parameters, such as network size $N$, intralayer interaction strength $A$ and interlayer coupling $K$.
The critical temperatures for the stability of disorder ($T_d$) and for the stability of order ($T_c$) are analogous to the critical temperatures observed in \cite{malarz_mean-field_2022}.

\begin{acknowledgments}
This research has received funding by the Polish National Science Center under Alphorn Grant No. 2019/01/Y/ST2/00058.
This work was funded by the European Union under the Horizon Europe grant OMINO (grant number 101086321). Views and opinions expressed are however those of the author(s) only and do not necessarily reflect those of the European Union or the European Research Executive Agency. Neither the European Union nor European Research Executive Agency can be held responsible for them.
This work has been co-funded by polish Ministry of Education and Science with the program International Project Co-Funding.
\end{acknowledgments}

\appendix
\section{Critical temperature calculations for duplex network}\label{App1}
Probabilities of the states $\vec{x}_{ij}$ having energy $E(\vec{x}_{ij})$ described in Eq.\ref{eq:energyxij} are according to the canonical ensemble (Eq. \ref{eq:canonical})
\begin{multline}
    P(\vec{x}_{ij})=\\
    \frac{e^{\left(\frac{-AM}{T}[ x_{ij}^{(1)} {(x^{(1)})}^2 + x_{ij}^{(2)} {(x^{(2)})}^2] \right)} 
    e^{\left( -\frac{K}{T} x_{ij}^{(1)}x_{ij}^{(2)}\right)}}{\sum_{\vec{x}_{mn}}e^{\left(\frac{-AM}{T}[ x_{mn}^{(1)} {(x^{(1)})}^2 + x_{mn}^{(2)} {(x^{(2)})}^2] \right)}
    e^{\left(-\frac{K}{T} x_{mn}^{(1)}x_{mn}^{(2)}\right)}} 
\end{multline}
where $M=N-2$ is the number of triads the link $ij$ is part of.
Then the expected value $\langle \vec{x}_{ij} \rangle$ (Eq.\ref{eq:meanx}) will be,
\begin{multline}
    \left\langle \vec{x}_{ij} \right\rangle =\\
    \sum_{\vec{x}_{ij}}  \frac{e^{\left(\frac{-AM}{T}[ x_{ij}^{(1)} {(x^{(1)})}^2 + x_{ij}^{(2)} {(x^{(2)})}^2] \right)} 
    e^{\left( -\frac{K}{T} x_{ij}^{(1)}x_{ij}^{(2)}\right)}}{\sum_{\vec{x}_{mn}}e^{\left(\frac{-AM}{T}[ x_{mn}^{(1)} {(x^{(1)})}^2 + x_{mn}^{(2)} {(x^{(2)})}^2] \right)}
    e^{\left(-\frac{K}{T} x_{mn}^{(1)}x_{mn}^{(2)}\right)}} \vec{x}_{ij} 
\end{multline}
Let us note that $\vec{x}_{ij}$ has only four possible values $\vec{x}_{ij} \in \{(-1,-1),(-1,1),(1,-1),(1,1)\}$, so both sums in the equation have limited number of terms.
Following mean-field methodology, we assume that the $\langle \vec{x}_{ij} \rangle$ calculated above is the same as the mean-field variable $\vec{x}$ through which the mean field is expressed, the above vector equation can be simplified into a set of self-consistent equations for $\vec{x}$
\begin{equation}
\begin{aligned}
x^{(1)}=f_1(x^{(1)},x^{(2)})\\
x^{(2)}=f_2(x^{(1)},x^{(2)})
\end{aligned}
\end{equation}
where the right-hand sides can be expressed by
\begin{multline}
f_{\alpha}(x^{(1)},x^{(2)})=\\
\frac{e^{2d}\sinh (a[{(x^{(1)})}^2\!\mathbb{+}{(x^{(2)})}^2]) \mathbb{+}\sinh (a(-1)^{\alpha}[{(x^{(2)})}^2\!\mathbb{-}{(x^{(1)})}^2])}{e^{2d}\cosh (a[{(x^{(1)})}^2+{(x^{(2)})}^2])+\cosh (a[{(x^{(1)})}^2-{(x^{(2)})}^2])}
\end{multline}
with $a=\frac{AM}{T}$, $d=\frac{K}{AM}$.
Note that the equations for both components differ only in the sign of the second hyperbolic sine function, expressed here as $(-1)^{\alpha}$.\\

\begin{figure}
    \centering
    \includegraphics[width=\linewidth]{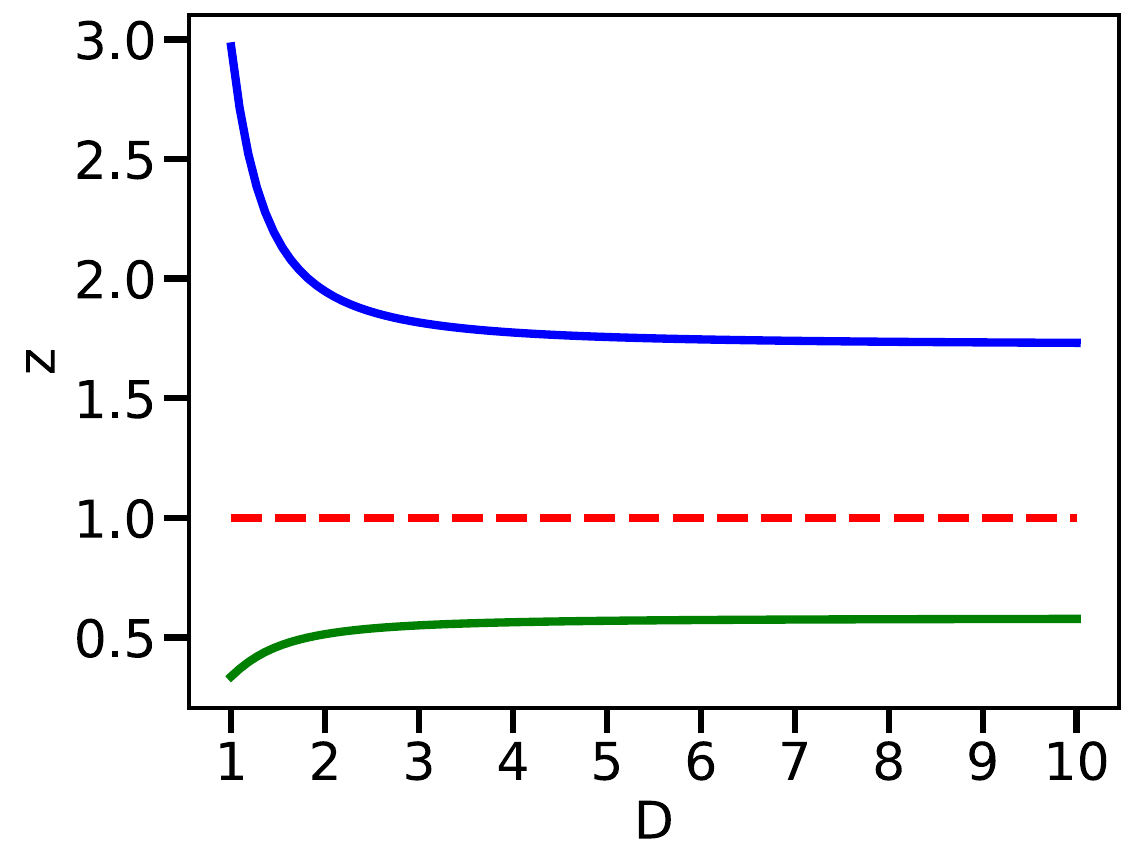} 
    \caption{Equation \ref{eq:trans1} has three solutions for auxiliary variable $z$ that depend only on relative interlayer coupling strength $D$. Out of them, only solution $z>1$ (blue) is physically meaningful and leads to a positive critical temperature $T_c$.}
    \label{fig:solutionz}
\end{figure}

The set of self-consistent equations above can be considered a map for the purposes of stability analysis, to determine whether a solution (fixed point) is actually stable and therefore a valid, stable state of the system it describes.
The Jacobian for the mean-field map (expressed also by Eq.\eqref{eq:2dmap}) is
\begin{equation}
    J= 
    \begin{bmatrix}
        \frac{\partial f_1}{\partial x^{(1)}} & 
        \frac{\partial f_1}{\partial x^{(2)}} \\[1ex]
        \frac{\partial f_2}{\partial x^{(1)}} & 
        \frac{\partial f_2}{\partial x^{(2)}} 
\end{bmatrix}
\end{equation}
with the derivatives equaling
\begin{equation}
\label{AEq:derivatives}
\begin{aligned}
    \frac{\partial f_1}{\partial x^{(1)}}&=\frac{4ax^{(1)}e^{2d}(\cosh(2d)+\cosh(2a{(x^{(2)})}^2)}{D}\\
    \frac{\partial f_1}{\partial x^{(2)}}&=\frac{2ax^{(2)}(e^{4d}-1)}{D}\\
    \frac{\partial f_2}{\partial x^{(1)}}&=\frac{2ax^{(1)}(e^{4d}-1)}{D}\\
    \frac{\partial f_2}{\partial x^{(2)}}&=\frac{4ax^{(2)}e^{2d}(\cosh(2d)+\cosh(2a{(x^{(1)})}^2)}{D}
\end{aligned}
\end{equation}
where
\begin{multline}
    \begin{aligned}
    D=\left(e^{2d}\cosh(a[{(x^{(1)})}^2+{(x^{(2)})}^2]) +\right.\\
    \left. \cosh(a[{(x^{(1)})}^2-{(x^{(2)})}^2])\right)^2
    \end{aligned}
\end{multline}
Since the two-dimensional maps are symmetric, to simplify calculations, we consider $x^{(1)}=x^{(2)}\equiv x$.
The fixed point of the mean-field map then becomes,
\begin{equation}\label{map2}
    x=\frac{\sinh(2ax^2)}{\cosh(2ax^2)+e^{-2d}}
\end{equation}
which in the limit of $d \rightarrow 0$ (no interactions) reduces to a single-layered network, as Eq. (9a) in \cite{malarz_mean-field_2022}, and for $d \rightarrow +\infty$ also reduces to the same equation, but with the interaction constant $2a$ instead of $a$.
The eigenvalues of the jacobian matrix $J$ given by Eq.(\ref{AEq:derivatives}) are,
\begin{equation}\label{eigen}
    \lambda_{+,-}=\frac{4ax(\cosh(2ax^2)+e^{\pm 2d})}{(e^d\left(\cosh(2ax^2)+e^{-2d}\right))^2}
\end{equation}
where $\lambda_+$ corresponds to the larger of the two eigenvalues. The critical temperature $T_c/(AM)$ and $x^c$ can be received from a pair of transcendental algebraic relations that describe the fixed point (Eq. (\ref{map2})) and criticality condition for the eigenvalue $\lambda_+=1$  (see E. (\ref{eigen})).
Multiplying Eq.(\ref{map2}) by inverse of Eq.(\ref{eigen}) with $\lambda_+=1$ allows us to combine the left-hand $x$ from (\ref{map2}) and $4ax$ term from (\ref{eigen}) into a single term $4ax^2$ which means that unknown $x$ only appears as $ax^2$.
Defining a new variable $z\equiv e^{ax^2}$ and noting $D\equiv e^d$, we can write a single equation for $z$
\begin{equation}\label{eq:trans1}
    8\ln z = \frac{(z^4-1)(z^4D^2+2z^2+D^2)}{z^2(z^4+2z^2D^2+1)}
\end{equation} 
This equation has 3 solutions in general, but only single relevant solution with $z>1$.
The solution, dependant on $D$ only, is shown in Fig. \ref{fig:solutionz}.
One can numerically calculate the largest solution $z$ for a fixed value of $d$.
Putting $ax^2=\ln z$ into Eq. (\ref{map2}) gives us the link polarization $x_c$ at the critical point
\begin{equation}
    x_c=\frac{D^2 (z^4-1)}{D^2(z^4+1)+2z^2}
\end{equation}
and having that, re-arranging definition of $z$ gives us the inverse of $a$ at the critical point
\begin{equation}\label{eq:criticalTempAppd}
    \frac{T_c}{AM}=\frac{x^2}{\ln z}
\end{equation}
To understand how the saturation critical temperature relates to a number of layers, we need to define the Eq.\ref{eq:criticalTempAppd} at $d\equiv D \to +\infty$ and at $d = 0$.
\begin{align}
    &\frac{T_c}{AM}\bigg|_{ d \to +\infty} = \frac{1}{\ln(z_{\infty})}\left(\frac{z_{\infty}^4-1}{z_{\infty}^4+1}\right)^2\\
    &\frac{T_c}{AM}\bigg|_{ d = 0} = \frac{1}{\ln(z_{0})}\left(\frac{(z_{0}^4-1)}{z_{0}^4+1+ 2z_{0}^2}\right)^2
\end{align}
Substituting the value of $z_{\infty}$=2.97 and $z_{0}$=8.823 obtained numerically from \ref{eq:trans1}, the saturation critical temperature is, 
\begin{equation}
    \frac{T_c\big|_{d\to +\infty}}{T_c\big|_{d= 0}}=\frac{1.16516\dots}{0.58258\dots}=2
\end{equation}
or 
\begin{equation}
    T_c\big|_{d\to +\infty} \approx 2 ~ T_c\big|_{d=0}
\end{equation}

\bibliography{main}

\providecommand{\noopsort}[1]{}\providecommand{\singleletter}[1]{#1}%
\begin{thebibliography}{35}%
\makeatletter
\providecommand \@ifxundefined [1]{%
 \@ifx{#1\undefined}
}%
\providecommand \@ifnum [1]{%
 \ifnum #1\expandafter \@firstoftwo
 \else \expandafter \@secondoftwo
 \fi
}%
\providecommand \@ifx [1]{%
 \ifx #1\expandafter \@firstoftwo
 \else \expandafter \@secondoftwo
 \fi
}%
\providecommand \natexlab [1]{#1}%
\providecommand \enquote  [1]{``#1''}%
\providecommand \bibnamefont  [1]{#1}%
\providecommand \bibfnamefont [1]{#1}%
\providecommand \citenamefont [1]{#1}%
\providecommand \href@noop [0]{\@secondoftwo}%
\providecommand \href [0]{\begingroup \@sanitize@url \@href}%
\providecommand \@href[1]{\@@startlink{#1}\@@href}%
\providecommand \@@href[1]{\endgroup#1\@@endlink}%
\providecommand \@sanitize@url [0]{\catcode `\\12\catcode `\$12\catcode
  `\&12\catcode `\#12\catcode `\^12\catcode `\_12\catcode `\%12\relax}%
\providecommand \@@startlink[1]{}%
\providecommand \@@endlink[0]{}%
\providecommand \url  [0]{\begingroup\@sanitize@url \@url }%
\providecommand \@url [1]{\endgroup\@href {#1}{\urlprefix }}%
\providecommand \urlprefix  [0]{URL }%
\providecommand \Eprint [0]{\href }%
\providecommand \doibase [0]{https://doi.org/}%
\providecommand \selectlanguage [0]{\@gobble}%
\providecommand \bibinfo  [0]{\@secondoftwo}%
\providecommand \bibfield  [0]{\@secondoftwo}%
\providecommand \translation [1]{[#1]}%
\providecommand \BibitemOpen [0]{}%
\providecommand \bibitemStop [0]{}%
\providecommand \bibitemNoStop [0]{.\EOS\space}%
\providecommand \EOS [0]{\spacefactor3000\relax}%
\providecommand \BibitemShut  [1]{\csname bibitem#1\endcsname}%
\let\auto@bib@innerbib\@empty
\bibitem [{\citenamefont {Kivel{\"a}}\ \emph {et~al.}(2014)\citenamefont
  {Kivel{\"a}}, \citenamefont {Arenas}, \citenamefont {Barthelemy},
  \citenamefont {Gleeson}, \citenamefont {Moreno},\ and\ \citenamefont
  {Porter}}]{kivela2014multilayer}%
  \BibitemOpen
  \bibfield  {author} {\bibinfo {author} {\bibfnamefont {M.}~\bibnamefont
  {Kivel{\"a}}}, \bibinfo {author} {\bibfnamefont {A.}~\bibnamefont {Arenas}},
  \bibinfo {author} {\bibfnamefont {M.}~\bibnamefont {Barthelemy}}, \bibinfo
  {author} {\bibfnamefont {J.~P.}\ \bibnamefont {Gleeson}}, \bibinfo {author}
  {\bibfnamefont {Y.}~\bibnamefont {Moreno}},\ and\ \bibinfo {author}
  {\bibfnamefont {M.~A.}\ \bibnamefont {Porter}},\ }\href
  {https://doi.org/10.1093/comnet/cnu016} {\bibfield  {journal} {\bibinfo
  {journal} {Journal of complex networks}\ }\textbf {\bibinfo {volume} {2}},\
  \bibinfo {pages} {203} (\bibinfo {year} {2014})}\BibitemShut {NoStop}%
\bibitem [{\citenamefont {Murase}\ \emph {et~al.}(2014)\citenamefont {Murase},
  \citenamefont {T\"or\"ok}, \citenamefont {Jo}, \citenamefont {Kaski},\ and\
  \citenamefont {Kert\'esz}}]{murase2014multilayer}%
  \BibitemOpen
  \bibfield  {author} {\bibinfo {author} {\bibfnamefont {Y.}~\bibnamefont
  {Murase}}, \bibinfo {author} {\bibfnamefont {J.}~\bibnamefont {T\"or\"ok}},
  \bibinfo {author} {\bibfnamefont {H.-H.}\ \bibnamefont {Jo}}, \bibinfo
  {author} {\bibfnamefont {K.}~\bibnamefont {Kaski}},\ and\ \bibinfo {author}
  {\bibfnamefont {J.}~\bibnamefont {Kert\'esz}},\ }\href
  {https://doi.org/10.1103/PhysRevE.90.052810} {\bibfield  {journal} {\bibinfo
  {journal} {Phys. Rev. E}\ }\textbf {\bibinfo {volume} {90}},\ \bibinfo
  {pages} {052810} (\bibinfo {year} {2014})}\BibitemShut {NoStop}%
\bibitem [{\citenamefont {Sol{\'e}-Ribalta}\ \emph {et~al.}(2016)\citenamefont
  {Sol{\'e}-Ribalta}, \citenamefont {G{\'o}mez},\ and\ \citenamefont
  {Arenas}}]{sole2016congestion}%
  \BibitemOpen
  \bibfield  {author} {\bibinfo {author} {\bibfnamefont {A.}~\bibnamefont
  {Sol{\'e}-Ribalta}}, \bibinfo {author} {\bibfnamefont {S.}~\bibnamefont
  {G{\'o}mez}},\ and\ \bibinfo {author} {\bibfnamefont {A.}~\bibnamefont
  {Arenas}},\ }\href {https://doi.org/10.1103/PhysRevLett.116.108701}
  {\bibfield  {journal} {\bibinfo  {journal} {Physical review letters}\
  }\textbf {\bibinfo {volume} {116}},\ \bibinfo {pages} {108701} (\bibinfo
  {year} {2016})}\BibitemShut {NoStop}%
\bibitem [{\citenamefont {Szell}\ \emph {et~al.}(2010)\citenamefont {Szell},
  \citenamefont {Lambiotte},\ and\ \citenamefont
  {Thurner}}]{szell2010multirelational}%
  \BibitemOpen
  \bibfield  {author} {\bibinfo {author} {\bibfnamefont {M.}~\bibnamefont
  {Szell}}, \bibinfo {author} {\bibfnamefont {R.}~\bibnamefont {Lambiotte}},\
  and\ \bibinfo {author} {\bibfnamefont {S.}~\bibnamefont {Thurner}},\ }\href
  {https://doi.org/10.1073/pnas.1004008107} {\bibfield  {journal} {\bibinfo
  {journal} {Proceedings of the National Academy of Sciences}\ }\textbf
  {\bibinfo {volume} {107}},\ \bibinfo {pages} {13636} (\bibinfo {year}
  {2010})}\BibitemShut {NoStop}%
\bibitem [{\citenamefont {Boccaletti}\ \emph {et~al.}(2014)\citenamefont
  {Boccaletti}, \citenamefont {Bianconi}, \citenamefont {Criado}, \citenamefont
  {Del~Genio}, \citenamefont {G{\'o}mez-Gardenes}, \citenamefont {Romance},
  \citenamefont {Sendina-Nadal}, \citenamefont {Wang},\ and\ \citenamefont
  {Zanin}}]{boccaletti2014structure}%
  \BibitemOpen
  \bibfield  {author} {\bibinfo {author} {\bibfnamefont {S.}~\bibnamefont
  {Boccaletti}}, \bibinfo {author} {\bibfnamefont {G.}~\bibnamefont
  {Bianconi}}, \bibinfo {author} {\bibfnamefont {R.}~\bibnamefont {Criado}},
  \bibinfo {author} {\bibfnamefont {C.~I.}\ \bibnamefont {Del~Genio}}, \bibinfo
  {author} {\bibfnamefont {J.}~\bibnamefont {G{\'o}mez-Gardenes}}, \bibinfo
  {author} {\bibfnamefont {M.}~\bibnamefont {Romance}}, \bibinfo {author}
  {\bibfnamefont {I.}~\bibnamefont {Sendina-Nadal}}, \bibinfo {author}
  {\bibfnamefont {Z.}~\bibnamefont {Wang}},\ and\ \bibinfo {author}
  {\bibfnamefont {M.}~\bibnamefont {Zanin}},\ }\href
  {https://doi.org/10.1016/j.physrep.2014.07.001} {\bibfield  {journal}
  {\bibinfo  {journal} {Physics reports}\ }\textbf {\bibinfo {volume} {544}},\
  \bibinfo {pages} {1} (\bibinfo {year} {2014})}\BibitemShut {NoStop}%
\bibitem [{\citenamefont {Hosni}\ \emph {et~al.}(2020)\citenamefont {Hosni},
  \citenamefont {Li},\ and\ \citenamefont {Ahmad}}]{hosni2020minimizing}%
  \BibitemOpen
  \bibfield  {author} {\bibinfo {author} {\bibfnamefont {A.~I.~E.}\
  \bibnamefont {Hosni}}, \bibinfo {author} {\bibfnamefont {K.}~\bibnamefont
  {Li}},\ and\ \bibinfo {author} {\bibfnamefont {S.}~\bibnamefont {Ahmad}},\
  }\href {https://doi.org/10.1016/j.ins.2019.10.063} {\bibfield  {journal}
  {\bibinfo  {journal} {Information Sciences}\ }\textbf {\bibinfo {volume}
  {512}},\ \bibinfo {pages} {1458} (\bibinfo {year} {2020})}\BibitemShut
  {NoStop}%
\bibitem [{\citenamefont {Singh}\ \emph {et~al.}(2008)\citenamefont {Singh},
  \citenamefont {Xu},\ and\ \citenamefont {Berger}}]{singh2008global}%
  \BibitemOpen
  \bibfield  {author} {\bibinfo {author} {\bibfnamefont {R.}~\bibnamefont
  {Singh}}, \bibinfo {author} {\bibfnamefont {J.}~\bibnamefont {Xu}},\ and\
  \bibinfo {author} {\bibfnamefont {B.}~\bibnamefont {Berger}},\ }\href
  {https://doi.org/10.1073/pnas.0806627105} {\bibfield  {journal} {\bibinfo
  {journal} {Proceedings of the National Academy of Sciences}\ }\textbf
  {\bibinfo {volume} {105}},\ \bibinfo {pages} {12763} (\bibinfo {year}
  {2008})}\BibitemShut {NoStop}%
\bibitem [{\citenamefont {G{\'o}mez-Gardenes}\ \emph
  {et~al.}(2012)\citenamefont {G{\'o}mez-Gardenes}, \citenamefont {Reinares},
  \citenamefont {Arenas},\ and\ \citenamefont
  {Flor{\'\i}a}}]{gomez2012evolution}%
  \BibitemOpen
  \bibfield  {author} {\bibinfo {author} {\bibfnamefont {J.}~\bibnamefont
  {G{\'o}mez-Gardenes}}, \bibinfo {author} {\bibfnamefont {I.}~\bibnamefont
  {Reinares}}, \bibinfo {author} {\bibfnamefont {A.}~\bibnamefont {Arenas}},\
  and\ \bibinfo {author} {\bibfnamefont {L.~M.}\ \bibnamefont {Flor{\'\i}a}},\
  }\href {https://doi.org/10.1038/srep00620} {\bibfield  {journal} {\bibinfo
  {journal} {Scientific reports}\ }\textbf {\bibinfo {volume} {2}},\ \bibinfo
  {pages} {620} (\bibinfo {year} {2012})}\BibitemShut {NoStop}%
\bibitem [{\citenamefont {Salehi}\ \emph {et~al.}(2015)\citenamefont {Salehi},
  \citenamefont {Sharma}, \citenamefont {Marzolla}, \citenamefont {Magnani},
  \citenamefont {Siyari},\ and\ \citenamefont {Montesi}}]{salehi2015spreading}%
  \BibitemOpen
  \bibfield  {author} {\bibinfo {author} {\bibfnamefont {M.}~\bibnamefont
  {Salehi}}, \bibinfo {author} {\bibfnamefont {R.}~\bibnamefont {Sharma}},
  \bibinfo {author} {\bibfnamefont {M.}~\bibnamefont {Marzolla}}, \bibinfo
  {author} {\bibfnamefont {M.}~\bibnamefont {Magnani}}, \bibinfo {author}
  {\bibfnamefont {P.}~\bibnamefont {Siyari}},\ and\ \bibinfo {author}
  {\bibfnamefont {D.}~\bibnamefont {Montesi}},\ }\href
  {https://doi.org/10.1109/TNSE.2015.2425961} {\bibfield  {journal} {\bibinfo
  {journal} {IEEE Transactions on Network Science and Engineering}\ }\textbf
  {\bibinfo {volume} {2}},\ \bibinfo {pages} {65 – 83} (\bibinfo {year}
  {2015})}\BibitemShut {NoStop}%
\bibitem [{\citenamefont {Arinik}\ \emph {et~al.}(2020)\citenamefont {Arinik},
  \citenamefont {Figueiredo},\ and\ \citenamefont
  {Labatut}}]{arinik2020partitioning}%
  \BibitemOpen
  \bibfield  {author} {\bibinfo {author} {\bibfnamefont {N.}~\bibnamefont
  {Arinik}}, \bibinfo {author} {\bibfnamefont {R.}~\bibnamefont {Figueiredo}},\
  and\ \bibinfo {author} {\bibfnamefont {V.}~\bibnamefont {Labatut}},\ }\href
  {https://doi.org/10.1016/j.socnet.2019.02.001} {\bibfield  {journal}
  {\bibinfo  {journal} {Social Networks}\ }\textbf {\bibinfo {volume} {60}},\
  \bibinfo {pages} {83 – 102} (\bibinfo {year} {2020})}\BibitemShut {NoStop}%
\bibitem [{\citenamefont {Buldyrev}\ \emph {et~al.}(2010)\citenamefont
  {Buldyrev}, \citenamefont {Parshani}, \citenamefont {Paul}, \citenamefont
  {Stanley},\ and\ \citenamefont {Havlin}}]{buldyrev2010cascading}%
  \BibitemOpen
  \bibfield  {author} {\bibinfo {author} {\bibfnamefont {S.~V.}\ \bibnamefont
  {Buldyrev}}, \bibinfo {author} {\bibfnamefont {R.}~\bibnamefont {Parshani}},
  \bibinfo {author} {\bibfnamefont {G.}~\bibnamefont {Paul}}, \bibinfo {author}
  {\bibfnamefont {H.~E.}\ \bibnamefont {Stanley}},\ and\ \bibinfo {author}
  {\bibfnamefont {S.}~\bibnamefont {Havlin}},\ }\href
  {https://doi.org/10.1038/nature08932} {\bibfield  {journal} {\bibinfo
  {journal} {Nature}\ }\textbf {\bibinfo {volume} {464}},\ \bibinfo {pages}
  {1025 – 1028} (\bibinfo {year} {2010})}\BibitemShut {NoStop}%
\bibitem [{\citenamefont {Atkisson}\ \emph {et~al.}(2020)\citenamefont
  {Atkisson}, \citenamefont {G{\'o}rski}, \citenamefont {Jackson},
  \citenamefont {Ho{\l}yst},\ and\ \citenamefont
  {D'Souza}}]{atkisson2020understanding}%
  \BibitemOpen
  \bibfield  {author} {\bibinfo {author} {\bibfnamefont {C.}~\bibnamefont
  {Atkisson}}, \bibinfo {author} {\bibfnamefont {P.~J.}\ \bibnamefont
  {G{\'o}rski}}, \bibinfo {author} {\bibfnamefont {M.~O.}\ \bibnamefont
  {Jackson}}, \bibinfo {author} {\bibfnamefont {J.~A.}\ \bibnamefont
  {Ho{\l}yst}},\ and\ \bibinfo {author} {\bibfnamefont {R.~M.}\ \bibnamefont
  {D'Souza}},\ }\href {https://doi.org/10.1002/evan.21850} {\bibfield
  {journal} {\bibinfo  {journal} {Evolutionary Anthropology: Issues, News, and
  Reviews}\ }\textbf {\bibinfo {volume} {29}},\ \bibinfo {pages} {102}
  (\bibinfo {year} {2020})}\BibitemShut {NoStop}%
\bibitem [{\citenamefont {Leskovec}\ \emph {et~al.}(2010)\citenamefont
  {Leskovec}, \citenamefont {Huttenlocher},\ and\ \citenamefont
  {Kleinberg}}]{leskovec2010signed}%
  \BibitemOpen
  \bibfield  {author} {\bibinfo {author} {\bibfnamefont {J.}~\bibnamefont
  {Leskovec}}, \bibinfo {author} {\bibfnamefont {D.}~\bibnamefont
  {Huttenlocher}},\ and\ \bibinfo {author} {\bibfnamefont {J.}~\bibnamefont
  {Kleinberg}},\ }in\ \href {https://doi.org/10.1145/1753326.1753532} {\emph
  {\bibinfo {booktitle} {Proceedings of the SIGCHI Conference on Human Factors
  in Computing Systems}}}\ (\bibinfo  {publisher} {Association for Computing
  Machinery},\ \bibinfo {address} {New York, NY, USA},\ \bibinfo {year}
  {2010})\ p.\ \bibinfo {pages} {1361–1370}\BibitemShut {NoStop}%
\bibitem [{\citenamefont {Heider}(1946)}]{heider1946attitudes}%
  \BibitemOpen
  \bibfield  {author} {\bibinfo {author} {\bibfnamefont {F.}~\bibnamefont
  {Heider}},\ }\href {https://doi.org/10.1080/00223980.1946.9917275} {\bibfield
   {journal} {\bibinfo  {journal} {The Journal of psychology}\ }\textbf
  {\bibinfo {volume} {21}},\ \bibinfo {pages} {107} (\bibinfo {year}
  {1946})}\BibitemShut {NoStop}%
\bibitem [{\citenamefont {Antal}\ \emph {et~al.}(2006)\citenamefont {Antal},
  \citenamefont {Krapivsky},\ and\ \citenamefont {Redner}}]{antal2006social}%
  \BibitemOpen
  \bibfield  {author} {\bibinfo {author} {\bibfnamefont {T.}~\bibnamefont
  {Antal}}, \bibinfo {author} {\bibfnamefont {P.~L.}\ \bibnamefont
  {Krapivsky}},\ and\ \bibinfo {author} {\bibfnamefont {S.}~\bibnamefont
  {Redner}},\ }\href {https://doi.org/10.1016/j.physd.2006.09.028} {\bibfield
  {journal} {\bibinfo  {journal} {Physica D: Nonlinear Phenomena}\ }\textbf
  {\bibinfo {volume} {224}},\ \bibinfo {pages} {130} (\bibinfo {year}
  {2006})}\BibitemShut {NoStop}%
\bibitem [{\citenamefont {Antal}\ \emph {et~al.}(2005)\citenamefont {Antal},
  \citenamefont {Krapivsky},\ and\ \citenamefont {Redner}}]{antal2005dynamics}%
  \BibitemOpen
  \bibfield  {author} {\bibinfo {author} {\bibfnamefont {T.}~\bibnamefont
  {Antal}}, \bibinfo {author} {\bibfnamefont {P.~L.}\ \bibnamefont
  {Krapivsky}},\ and\ \bibinfo {author} {\bibfnamefont {S.}~\bibnamefont
  {Redner}},\ }\href {https://doi.org/10.1103/PhysRevE.72.036121} {\bibfield
  {journal} {\bibinfo  {journal} {Physical Review E}\ }\textbf {\bibinfo
  {volume} {72}},\ \bibinfo {pages} {036121} (\bibinfo {year}
  {2005})}\BibitemShut {NoStop}%
\bibitem [{\citenamefont {Srinivasan}(2011)}]{srinivasan2011local}%
  \BibitemOpen
  \bibfield  {author} {\bibinfo {author} {\bibfnamefont {A.}~\bibnamefont
  {Srinivasan}},\ }\href {https://doi.org/10.1073/pnas.1018901108} {\bibfield
  {journal} {\bibinfo  {journal} {Proceedings of the National Academy of
  Sciences}\ }\textbf {\bibinfo {volume} {108}},\ \bibinfo {pages} {1751}
  (\bibinfo {year} {2011})}\BibitemShut {NoStop}%
\bibitem [{\citenamefont {Cartwright}\ and\ \citenamefont
  {Harary}(1956)}]{cartwright1956structural}%
  \BibitemOpen
  \bibfield  {author} {\bibinfo {author} {\bibfnamefont {D.}~\bibnamefont
  {Cartwright}}\ and\ \bibinfo {author} {\bibfnamefont {F.}~\bibnamefont
  {Harary}},\ }\href@noop {} {\bibfield  {journal} {\bibinfo  {journal}
  {Psychological review}\ }\textbf {\bibinfo {volume} {63}},\ \bibinfo {pages}
  {277} (\bibinfo {year} {1956})}\BibitemShut {NoStop}%
\bibitem [{\citenamefont {Marvel}\ \emph {et~al.}(2011)\citenamefont {Marvel},
  \citenamefont {Kleinberg}, \citenamefont {Kleinberg},\ and\ \citenamefont
  {Strogatz}}]{marvel2011continuous}%
  \BibitemOpen
  \bibfield  {author} {\bibinfo {author} {\bibfnamefont {S.~A.}\ \bibnamefont
  {Marvel}}, \bibinfo {author} {\bibfnamefont {J.}~\bibnamefont {Kleinberg}},
  \bibinfo {author} {\bibfnamefont {R.~D.}\ \bibnamefont {Kleinberg}},\ and\
  \bibinfo {author} {\bibfnamefont {S.~H.}\ \bibnamefont {Strogatz}},\ }\href
  {https://doi.org/10.1073/pnas.1013213108} {\bibfield  {journal} {\bibinfo
  {journal} {Proceedings of the National Academy of Sciences of the United
  States of America}\ }\textbf {\bibinfo {volume} {108}},\ \bibinfo {pages}
  {1771 – 1776} (\bibinfo {year} {2011})}\BibitemShut {NoStop}%
\bibitem [{\citenamefont {Rabbani}\ \emph {et~al.}(2019)\citenamefont
  {Rabbani}, \citenamefont {Shirazi},\ and\ \citenamefont
  {Jafari}}]{rabbani2019mean}%
  \BibitemOpen
  \bibfield  {author} {\bibinfo {author} {\bibfnamefont {F.}~\bibnamefont
  {Rabbani}}, \bibinfo {author} {\bibfnamefont {A.~H.}\ \bibnamefont
  {Shirazi}},\ and\ \bibinfo {author} {\bibfnamefont {G.}~\bibnamefont
  {Jafari}},\ }\href {https://doi.org/10.1103/PhysRevE.99.062302} {\bibfield
  {journal} {\bibinfo  {journal} {Physical Review E}\ }\textbf {\bibinfo
  {volume} {99}},\ \bibinfo {pages} {062302} (\bibinfo {year}
  {2019})}\BibitemShut {NoStop}%
\bibitem [{\citenamefont {Malarz}\ and\ \citenamefont
  {Ku\l{}akowski}(2021)}]{malarz2021comment}%
  \BibitemOpen
  \bibfield  {author} {\bibinfo {author} {\bibfnamefont {K.}~\bibnamefont
  {Malarz}}\ and\ \bibinfo {author} {\bibfnamefont {K.}~\bibnamefont
  {Ku\l{}akowski}},\ }\href {https://doi.org/10.1103/PhysRevE.103.066301}
  {\bibfield  {journal} {\bibinfo  {journal} {Phys. Rev. E}\ }\textbf {\bibinfo
  {volume} {103}},\ \bibinfo {pages} {066301} (\bibinfo {year}
  {2021})}\BibitemShut {NoStop}%
\bibitem [{\citenamefont {Shojaei}\ \emph {et~al.}(2019)\citenamefont
  {Shojaei}, \citenamefont {Manshour},\ and\ \citenamefont
  {Montakhab}}]{shojaei2019phase}%
  \BibitemOpen
  \bibfield  {author} {\bibinfo {author} {\bibfnamefont {R.}~\bibnamefont
  {Shojaei}}, \bibinfo {author} {\bibfnamefont {P.}~\bibnamefont {Manshour}},\
  and\ \bibinfo {author} {\bibfnamefont {A.}~\bibnamefont {Montakhab}},\ }\href
  {https://doi.org/10.1103/PhysRevE.100.022303} {\bibfield  {journal} {\bibinfo
   {journal} {Phys. Rev. E}\ }\textbf {\bibinfo {volume} {100}},\ \bibinfo
  {pages} {022303} (\bibinfo {year} {2019})}\BibitemShut {NoStop}%
\bibitem [{\citenamefont {Summers}\ and\ \citenamefont
  {Shames}(2013)}]{summers2013active}%
  \BibitemOpen
  \bibfield  {author} {\bibinfo {author} {\bibfnamefont {T.~H.}\ \bibnamefont
  {Summers}}\ and\ \bibinfo {author} {\bibfnamefont {I.}~\bibnamefont
  {Shames}},\ }\href {https://doi.org/10.1209/0295-5075/103/18001} {\bibfield
  {journal} {\bibinfo  {journal} {Europhysics Letters}\ }\textbf {\bibinfo
  {volume} {103}},\ \bibinfo {pages} {18001} (\bibinfo {year}
  {2013})}\BibitemShut {NoStop}%
\bibitem [{\citenamefont {Xia}\ \emph {et~al.}(2015)\citenamefont {Xia},
  \citenamefont {Cao},\ and\ \citenamefont {Johansson}}]{xia2015structural}%
  \BibitemOpen
  \bibfield  {author} {\bibinfo {author} {\bibfnamefont {W.}~\bibnamefont
  {Xia}}, \bibinfo {author} {\bibfnamefont {M.}~\bibnamefont {Cao}},\ and\
  \bibinfo {author} {\bibfnamefont {K.~H.}\ \bibnamefont {Johansson}},\ }\href
  {https://doi.org/10.1109/TCNS.2015.2437528} {\bibfield  {journal} {\bibinfo
  {journal} {IEEE Transactions on Control of Network Systems}\ }\textbf
  {\bibinfo {volume} {3}},\ \bibinfo {pages} {46} (\bibinfo {year}
  {2015})}\BibitemShut {NoStop}%
\bibitem [{\citenamefont {Yang}\ \emph {et~al.}(2022)\citenamefont {Yang},
  \citenamefont {Wang}, \citenamefont {Ma}, \citenamefont {He},\ and\
  \citenamefont {Huang}}]{yang2022promotive}%
  \BibitemOpen
  \bibfield  {author} {\bibinfo {author} {\bibfnamefont {M.}~\bibnamefont
  {Yang}}, \bibinfo {author} {\bibfnamefont {X.}~\bibnamefont {Wang}}, \bibinfo
  {author} {\bibfnamefont {L.}~\bibnamefont {Ma}}, \bibinfo {author}
  {\bibfnamefont {Q.}~\bibnamefont {He}},\ and\ \bibinfo {author}
  {\bibfnamefont {M.}~\bibnamefont {Huang}},\ }\href
  {https://doi.org/10.1007/s00521-022-07298-y} {\bibfield  {journal} {\bibinfo
  {journal} {Neural Computing and Applications}\ }\textbf {\bibinfo {volume}
  {34}},\ \bibinfo {pages} {16683} (\bibinfo {year} {2022})}\BibitemShut
  {NoStop}%
\bibitem [{\citenamefont {Facchetti}\ \emph {et~al.}(2011)\citenamefont
  {Facchetti}, \citenamefont {Iacono},\ and\ \citenamefont
  {Altafini}}]{facchetti2011computing}%
  \BibitemOpen
  \bibfield  {author} {\bibinfo {author} {\bibfnamefont {G.}~\bibnamefont
  {Facchetti}}, \bibinfo {author} {\bibfnamefont {G.}~\bibnamefont {Iacono}},\
  and\ \bibinfo {author} {\bibfnamefont {C.}~\bibnamefont {Altafini}},\ }\href
  {https://doi.org/10.1073/pnas.1109521108} {\bibfield  {journal} {\bibinfo
  {journal} {Proceedings of the National Academy of Sciences}\ }\textbf
  {\bibinfo {volume} {108}},\ \bibinfo {pages} {20953} (\bibinfo {year}
  {2011})}\BibitemShut {NoStop}%
\bibitem [{\citenamefont {Doreian}\ and\ \citenamefont
  {Mrvar}(2015)}]{doreian2015structural}%
  \BibitemOpen
  \bibfield  {author} {\bibinfo {author} {\bibfnamefont {P.}~\bibnamefont
  {Doreian}}\ and\ \bibinfo {author} {\bibfnamefont {A.}~\bibnamefont
  {Mrvar}},\ }\href {https://doi.org/10.21307/joss-2019-012} {\bibfield
  {journal} {\bibinfo  {journal} {Journal of Social Structure}\ }\textbf
  {\bibinfo {volume} {16}},\ \bibinfo {pages} {1} (\bibinfo {year}
  {2015})}\BibitemShut {NoStop}%
\bibitem [{\citenamefont {Górski}\ \emph {et~al.}(2017)\citenamefont
  {Górski}, \citenamefont {Kułakowski}, \citenamefont {Gawroński},\ and\
  \citenamefont {Hołyst}}]{gorski_destructive_2017}%
  \BibitemOpen
  \bibfield  {author} {\bibinfo {author} {\bibfnamefont {P.~J.}\ \bibnamefont
  {Górski}}, \bibinfo {author} {\bibfnamefont {K.}~\bibnamefont
  {Kułakowski}}, \bibinfo {author} {\bibfnamefont {P.}~\bibnamefont
  {Gawroński}},\ and\ \bibinfo {author} {\bibfnamefont {J.~A.}\ \bibnamefont
  {Hołyst}},\ }\href {https://doi.org/10.1038/s41598-017-15960-y} {\bibfield
  {journal} {\bibinfo  {journal} {Scientific Reports}\ }\textbf {\bibinfo
  {volume} {7}},\ \bibinfo {pages} {16047} (\bibinfo {year}
  {2017})}\BibitemShut {NoStop}%
\bibitem [{\citenamefont {Kundu}\ and\ \citenamefont
  {Pandey}(2022)}]{kundu2022balance}%
  \BibitemOpen
  \bibfield  {author} {\bibinfo {author} {\bibfnamefont {R.~P.}\ \bibnamefont
  {Kundu}}\ and\ \bibinfo {author} {\bibfnamefont {S.}~\bibnamefont {Pandey}},\
  }\href {https://papers.ssrn.com/abstract=4174703} {\bibfield  {journal}
  {\bibinfo  {journal} {Available at SSRN 4174703}\ } (\bibinfo {year}
  {2022})}\BibitemShut {NoStop}%
\bibitem [{\citenamefont {Cozzo}\ \emph {et~al.}(2015)\citenamefont {Cozzo},
  \citenamefont {Kivel{\"a}}, \citenamefont {De~Domenico}, \citenamefont
  {Sol{\'e}-Ribalta}, \citenamefont {Arenas}, \citenamefont {G{\'o}mez},
  \citenamefont {Porter},\ and\ \citenamefont {Moreno}}]{cozzo2015structure}%
  \BibitemOpen
  \bibfield  {author} {\bibinfo {author} {\bibfnamefont {E.}~\bibnamefont
  {Cozzo}}, \bibinfo {author} {\bibfnamefont {M.}~\bibnamefont {Kivel{\"a}}},
  \bibinfo {author} {\bibfnamefont {M.}~\bibnamefont {De~Domenico}}, \bibinfo
  {author} {\bibfnamefont {A.}~\bibnamefont {Sol{\'e}-Ribalta}}, \bibinfo
  {author} {\bibfnamefont {A.}~\bibnamefont {Arenas}}, \bibinfo {author}
  {\bibfnamefont {S.}~\bibnamefont {G{\'o}mez}}, \bibinfo {author}
  {\bibfnamefont {M.~A.}\ \bibnamefont {Porter}},\ and\ \bibinfo {author}
  {\bibfnamefont {Y.}~\bibnamefont {Moreno}},\ }\href
  {https://doi.org/10.1088/1367-2630/17/7/073029} {\bibfield  {journal}
  {\bibinfo  {journal} {New Journal of Physics}\ }\textbf {\bibinfo {volume}
  {17}},\ \bibinfo {pages} {073029} (\bibinfo {year} {2015})}\BibitemShut
  {NoStop}%
\bibitem [{\citenamefont {Burghardt}\ and\ \citenamefont
  {Maoz}(2018)}]{burghardt2018imbalance}%
  \BibitemOpen
  \bibfield  {author} {\bibinfo {author} {\bibfnamefont {K.}~\bibnamefont
  {Burghardt}}\ and\ \bibinfo {author} {\bibfnamefont {Z.}~\bibnamefont
  {Maoz}},\ }\bibfield  {journal} {\bibinfo  {journal} {Available at SSRN
  3192890}\ }\href {https://doi.org/10.2139/ssrn.3192890}
  {10.2139/ssrn.3192890} (\bibinfo {year} {2018})\BibitemShut {NoStop}%
\bibitem [{\citenamefont {Ku{\l}akowski}\ \emph {et~al.}(2005)\citenamefont
  {Ku{\l}akowski}, \citenamefont {Gawro{\'n}ski},\ and\ \citenamefont
  {Gronek}}]{kulakowski2005heider}%
  \BibitemOpen
  \bibfield  {author} {\bibinfo {author} {\bibfnamefont {K.}~\bibnamefont
  {Ku{\l}akowski}}, \bibinfo {author} {\bibfnamefont {P.}~\bibnamefont
  {Gawro{\'n}ski}},\ and\ \bibinfo {author} {\bibfnamefont {P.}~\bibnamefont
  {Gronek}},\ }\href {https://doi.org/10.1142/S012918310500742X} {\bibfield
  {journal} {\bibinfo  {journal} {International Journal of Modern Physics C}\
  }\textbf {\bibinfo {volume} {16}},\ \bibinfo {pages} {707} (\bibinfo {year}
  {2005})}\BibitemShut {NoStop}%
\bibitem [{\citenamefont {Malarz}\ and\ \citenamefont
  {Hołyst}(2022)}]{malarz_mean-field_2022}%
  \BibitemOpen
  \bibfield  {author} {\bibinfo {author} {\bibfnamefont {K.}~\bibnamefont
  {Malarz}}\ and\ \bibinfo {author} {\bibfnamefont {J.~A.}\ \bibnamefont
  {Hołyst}},\ }\href {https://doi.org/10.1103/PhysRevE.106.064139} {\bibfield
  {journal} {\bibinfo  {journal} {Physical Review E}\ }\textbf {\bibinfo
  {volume} {106}},\ \bibinfo {pages} {064139} (\bibinfo {year}
  {2022})}\BibitemShut {NoStop}%
\bibitem [{\citenamefont {Helbing}\ \emph {et~al.}(2000)\citenamefont
  {Helbing}, \citenamefont {Farkas},\ and\ \citenamefont
  {Vicsek}}]{helbing2000freezing}%
  \BibitemOpen
  \bibfield  {author} {\bibinfo {author} {\bibfnamefont {D.}~\bibnamefont
  {Helbing}}, \bibinfo {author} {\bibfnamefont {I.~J.}\ \bibnamefont
  {Farkas}},\ and\ \bibinfo {author} {\bibfnamefont {T.}~\bibnamefont
  {Vicsek}},\ }\href {https://doi.org/10.1103/PhysRevLett.84.1240} {\bibfield
  {journal} {\bibinfo  {journal} {Physical Review Letters}\ }\textbf {\bibinfo
  {volume} {84}},\ \bibinfo {pages} {1240 – 1243} (\bibinfo {year}
  {2000})}\BibitemShut {NoStop}%
\bibitem [{\citenamefont {Hołyst}(2023)}]{holyst2023history}%
  \BibitemOpen
  \bibfield  {author} {\bibinfo {author} {\bibfnamefont {J.~A.}\ \bibnamefont
  {Hołyst}},\ }\href
  {https://doi.org/https://doi.org/10.1016/j.jocs.2023.102137} {\bibfield
  {journal} {\bibinfo  {journal} {Journal of Computational Science}\ }\textbf
  {\bibinfo {volume} {73}},\ \bibinfo {pages} {102137} (\bibinfo {year}
  {2023})}\BibitemShut {NoStop}%
\end{thebibliography}%

\end{document}